\documentclass[twocolumn]{revtex4-1}
\usepackage[margin=1in]{geometry}
\usepackage[parfill]{parskip}
\usepackage{graphicx}
\usepackage{amssymb}
\usepackage{epstopdf}
\usepackage{amsmath}
\usepackage{hyperref}
\usepackage{multirow}
\usepackage{caption}
\usepackage{subcaption}
\usepackage{float}
\usepackage{titlesec}
\usepackage{lipsum}
\usepackage[utf8]{inputenc}
\usepackage[usenames]{xcolor}

\def\be{\begin{equation}}
\def\ee{\end{equation}}

\def\bea{\begin{eqnarray}}
\def\eea{\end{eqnarray}}

\def\bci{\begin{compactitem}}
\def\eci{\end{compactitem}}

\def\la{\label}
\def\non{\nonumber}
\def\fr{\frac}
\def\ci{\cite}
\def\le{\left}
\def\ri{\right}
\def\eq{\equiv}

\def\rbao{r_{BAO}}
\def\a{\alpha}
\def\Ode{\Omega_{DE}}
\def\Om{\Omega_m}
\def\Or{\Omega_{r}}

\def\rde{\rho_{de}}

\def\La{\Lambda}
\def\s{\sigma}

\begin{document}


\title{Constraints on Steep Equation of State for the Dark Energy using BAO}
\author{Mariana Jaber and Axel de la Macorra }
\affiliation{Instituto de Fisica, Universidad Nacional Autonoma de Mexico,\\  A.P. 20-364, 01000, Mexico D.F., Mexico}



\begin{abstract}

We present a parametrization for the Dark Energy Equation of State ``EoS"  which has a rich structure. Our EoS has a transition at pivotal redshift $z_T$ between the  present day value $w_0$ to  an early time $w_i=w_a+w_0\equiv w(z>>0)$ and the steepness of this transition is given in terms of the $q$ parameter. The proposed parametrization is  $w=w_0+w_a(z/z_T)^q/(1+(z/z_T))^q$, with $w_0$, $w_i$, $q $ and $z_T$ constant parameters. This transition is motivated by scalar field  dynamics such as for example quintessence models. Our parametrization reduces to  the widely used EoS $w=w_0+w_a(1-a)$ for $z_T=q=1$. We study if a late time transition is favored by BAO measurements and Planck priors. According to our results, an EoS with a present value of $w_0 = -0.91$ and a high redshifts value $w_i =-0.62$, featuring a transition at a redshift of $z_T =  1.16$ with an exponent $q = 9.95$ is a good fit to the observational data. We found good agreement between the model and the data reported by the different surveys. A ``thawing`" dynamics is preferred by the use of BAO data alone (including Lymman-$\alpha$ forest measurements) and a ``freezing" evolution of the EoS is preferred when we include the priors from Planck.
The constraints imposed by the available BAO measurements (\ci{Beutler:2011hx, Ross:2014qpa, Anderson:2013oza, Kazin:2014qga, Font-Ribera:2013wce, Delubac:2014aqe, Gong:2015tta}) and its physical behavior are discussed.  

\end{abstract}

\maketitle


\section{Introduction}
 
We are in a very particular epoch of the cosmic history where the expansion of the Universe is accelerated due to an unknown energy component commonly referred to as Dark Energy (DE).  There is strong evidence that supports an accelerated expansion coming from observations of Type Ia supernovae (SNIa) (Riess {\it et al.} 1998 \ci{Riess:1998cb} and Perl\-mu\-tter {\it et al.} 1999 \ci{Perlmutter:1998np}), cosmic microwave background (CMB) (WMAP Collaboration \ci{Bennett:2012zja}, Planck Co\-lla\-bo\-ra\-tion \ci{Ade:2015xua}), large scale structure (LSS) (Tegmark {\it et al.} \ci{Tegmark:2003ud}, Cole {\it et al. } 2005 \ci{Cole:2005sx}) and baryon acoustic oscillations (\ci{Eisenstein:2005su, Colless:2003wz, Gong:2015tta, Beutler:2011hx, Kazin:2014qga, Anderson:2013oza, Ross:2014qpa}). According to the observations our Universe is flat and dominated at present time by this DE component.

Among the proposals, models to parametrize the DE equation of state (EoS) as a function of redshift have arisen in the literature (\ci{Chevallier:2000qy, Linder:2002et, Doran:2006kp, Linder:2006ud, Rubin:2008wq, 2009ApJ...703.1374S, 2010PhRvD..81f3007M, Hannestad:2004cb, Jassal:2004ej, Ma:2011nc, Huterer:2000mj, Weller:2001gf, Huang:2010zra, delaMacorra:2015aqf}). The most po\-pu\-lar among them is so called CPL parametrization, given by $w=w_0+w_a(1-a)$,  widely used in many cosmological observational analysis. The  value of the equation of state of DE is restricted by observations to be close to $-1$  ($ w = -1.019^{+0.075}_{-0.080}$ according to the 95$\%$ limits imposed by \emph{Planck } data combined with  other astrophysical measurements  \ci{Ade:2015xua}).
Nevertheless, the behavior and properties at different cosmic epochs is much poorly constrained by current observations.
Therefore we are interested in studying $w$ at a late time and see if a transition in the EoS takes place.  The  parametrization used here is $w=w_0+w_a(z/z_T)^q/(1+(z/z_T))^q$, with $w_0$, $w_a=w_i-w_0$, $q$ and $z_T$ for  constant parameters. This EoS allows for a steep transition for a large value of $q$ and the  pivotal point is $z_T$ with $w(z_T)=(w_0+w_i)/2$ giving the middle point of the transition between $w_0$  and  the early time value $w_i=w(z\gg 1)$. This transition is motivated by scalar field  dynamics such as quintessence models \ci{delaMacorra:1999ff} and  in \ci{delaMacorra:2015aqf} a new parametrization that  encapsules the dynamics of DE was presented. However, in this work we use simpler structure for the EoS since we are interested in determining  the late time behavior of Dark Energy using BAO measurements with Planck priors. 

Perhaps the best physically motivated candidates for Dark Energy are scalar fields with only gravitational interaction  \ci{Ratra:1987rm,Wetterich:1994bg, PhysRevD.59.123504, delaMacorra:1999ff} and special interest was devoted to tracker fields \ci{PhysRevD.59.123504}, since in this case the behavior of the scalar field $\phi$ is weakly dependent on the initial conditions set at an early epoch, well before matter-radiation equality. In this class of models a fundamental question is why DE is relevant now, know as the coincidence problem, and this can be understood by the insensitivity of the late time dynamics on the initial conditions of $\phi$. However,  tracker fields may not give the correct phenomenology since they have a large value of $w$ at present time.  We are more interested at this stage  to work from present day redshift  $z=0$ to larger values of $z$, in the region where DE is most relevant. Interesting models for DE and DM have been proposed using gauge groups, similar to QCD in particle physics, and have been studied to understand the nature of Dark Energy \ci{delaMacorra:2004mp, DelaMacorra:2001uq,delaMacorra:2001ay} and also Dark Matter \ci{2010APh....33..195D, delaMacorra:2002tk}.

The scientific community is devoting a large amount of time and resources to investigate the dynamics and nature of Dark Energy, working on current (SDSS-IV \cite{Dawson:2015wdb}, DES \cite{Abbott:2005bi}) and future  (DESI \cite{Levi:2013gra}, Euclid \cite{2011arXiv1110.3193L}, LSST \cite{2009arXiv0912.0201L}) experiments to study with  very high precision the expansion history of the Universe and test interesting models beyond a cosmological constant or Taylor expansions of the equation of state of Dark Energy.

This article is organized as follows: we introduce our basic cosmological framework  in Section \ref{sect:Method}, Section \ref{sec:Analysis} details the analysis performed, the  results obtained are in Section \ref{sec:Results} while Section \ref{sec:Conclusions} summarizes our Conclusions. 


\section{Methodology}
	\label{sect:Method}

\subsection{Cosmology}
	\label{subsect:Cosmo}	

We assume the validity of General Relativity and work within a flat Universe in a FRW metric.  The Friedman equation can thus be expressed in terms of the redshift $z$ as

\be
	\la{eq:Hz}
	H(z) = H_0\sqrt{\Or(1+z)^4+\Om(1+z)^3+\Ode F(z)}
\ee

where $H\eq(\fr{da}{dt})(\fr{1}{a})$ is the Hubble parameter, $a=(1+z)^{-1}$ the scale factor of the Universe and  $H_0 = 100\cdot h$ $km\cdot s^{-1} Mpc^{-1}$ is the Hubble constant at present time. We are using the standard definition  $a_0 =1$ and  $t$ to represent the cosmic time. The present fractional densities of matter, radiation and Dark Energy (DE) are given by $\Om$, $\Or$, $\Ode$, respectively. The function $F(z)$ in equation  \ref{eq:Hz} encodes the evolution of the DE component in terms of its equation of state (EoS), $w(z)$,

\bea
	\la{eq:DE-f}
	F(z) &\eq& \fr{\rde(z)}{\rde(0)} \\ \non
	F(z) &=& exp\le( -3 \int_0^z dz'\fr{1+w(z')}{1+z'}\ri)
\eea

The EoS, $w(z)$, introduced in equation \ref{eq:DE-f}  specifies the evolution of our DE fluid and accordingly, the rate of expansion  of the Universe at late times, following the dynamics set by Equation \ref{eq:Hz}.

\subsection{Dark Energy  Equation of State}
	\label{subsect:DEeos}

We are concerned with the study of DE only at background level for this work and we model  the DE equation of state (EoS) with the following parametrization:

\bea
	\la{eq:eos}
	w_{de}(z)& = & w_0 + (w_i - w_0) f(z) \non \\
	& = & w_0 + (w_i - w_0) \fr{(\fr{z}{z_T})^q}{1+(\fr{z}{z_T})^q}, \\
	f(z) &\eq & \fr{(\fr{z}{z_T})^q}{1+(\fr{z}{z_T})^q} \non
\eea

where $w_i$ and $w_0$ represent the value for $w(z)$ at large redshifts and at present day, respectively whereas the function $f(z)$ modulates the dynamics of this parametrization in between both values, and takes the values $f(z=0)=0$, $f(z\rightarrow \infty)=1$ and $f(z=z_t)=1/2$, \emph{i.e.} $0\leq f(z) \leq 1$.

One thing to be noted is that this EoS does not have a constant slope within the two regimes: $w(z=0) \rightarrow w_0$, $w(z>>0) \rightarrow w_i$, but it makes a transition between them at a redshift $z = z_t$, taking a value of $w(z_t) = \fr{w_0+w_i}{2}$. The parameter $q$ modulates the steepness of the transition featured, as shown in Figure \ref{fig:eos}.

\begin{figure}[t]
	\centering
		\includegraphics[width=\linewidth]{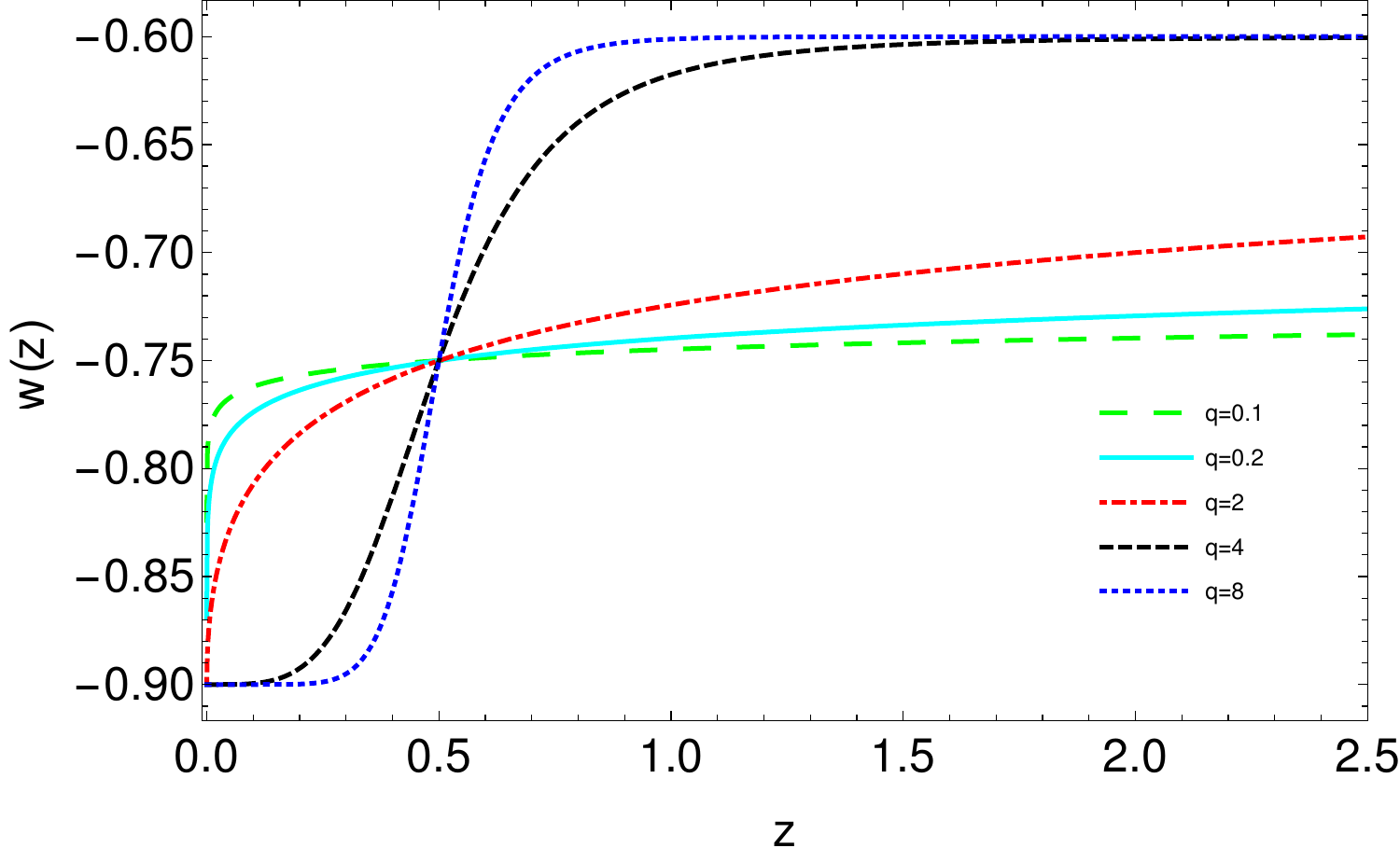}
		\caption{Evolution of the EoS according to equation \ref{eq:eos} for fixed values  $w_0 = -0.9$, $w_1=-0.7$, $z_T=0.5$. The green and cyan curves correspond to EoS with $q<1$ and the remaining three curves represent a EoS with $q>1$. }
		\la{fig:eos}
\end{figure}

In particular, for $q \geq 1$ we can see a direct connection  between the value of $q$ and the steepness of the transition: the greater the value for $q$, the steeper the transition we will have. For $q<1$ we have a very different behavior. As $q \rightarrow 0$, $w(z)$ performs an increasingly rapid transition from $w_0$ to $\approx(w_0+w_i)/2$ at $z\approx 0$ and continues to evolve with a nearly constant behavior regardless of the $z_t$ value, as seen in Figure \ref{fig:eos}. This is obvious since  $\lim_{q\to0}\fr{(\fr{z}{z_T})^q}{1+(\fr{z}{z_T})^q}=1/2$ and so $\lim_{q\to0}w(z)=(w_0+w_i)/2$.
The parametrization in equation \ref{eq:eos} includes the usual CPL pa\-ra\-me\-tri\-za\-tion (\ci{Chevallier:2000qy, Linder:2002et}) 
\be
	\la{eq:CPL}
	w(z)\vert_{q=1=z_T}=  w_0 + w_a\fr{z}{1+z} 
\ee
as a particular case when  $q = 1$ and $z_t =1$  but  it allows for a richer physical behavior.

Clearly, taking $w_0 = w_i = -1$, the cosmological cons\-tant solution is recovered.

\subsection{BAO as a cosmological probe}

	\la{subsect:BAO}

Considering that we want to test the dynamics of Dark Energy we are interested in the expansion history of the Universe at late times when  it is driven by the DE component. Ever since its first detection (Cole et al. 2005 \ci{Colless:2003wz}, Eisenstein et al 2005, \ci{Eisenstein:2005su}) the Baryon Acoustic Osci\-lla\-tion (BAO) feature has been widely used as a po\-wer\-ful probe for cosmology becoming the standard rulers of choice just as Type Ia supernovae (SNIa) were at the early part of the 21st century during the beginning of the so called  ``distance revolution". 

The BAO feature has become the best way to probe the late time dynamics of the Universe and in consequence that of  DE. It is the cosmological tool used by several experiments like the SDSS-IV \cite{Dawson:2015wdb} and the Dark Energy Survey (DES) \cite{Abbott:2005bi} and the main  probe that will be used by future experiments like the Dark Energy Spectroscopic Instrument (DESI) \cite{Levi:2013gra} and Euclid \cite{2011arXiv1110.3193L}. Nevertheless a complete analysis should rely on the data provided by the recombination era, since the CMB provides the most accurate constraints on the cosmological parameters. 

This standard ruler is set by a particular size in the spatial distribution of matter which can be used to constrain the  parameters in equation  \ref{eq:eos}. The co\-rres\-pon\-ding size, $\rbao(z)$, is obtained by performing a spherical average of the distribution of galaxies both along and across the line of sight (Bassett and  Hlozek 2010 \ci{Bassett:2009mm})

\be
	\la{eq:rbao}
	\rbao(z) \eq \frac{s(z_d)}{D_V(z)}
\ee

The comoving sound horizon at the baryon drag epoch is represented by $s_d$  and the dilation scale, $D_V(z)$, contains the information about the cos\-mo\-lo\-gy used in $H(z)$:

\bea
	\la{eq:sd}
	s(z_d) & \eq &  \fr{c}{H_0} \int_{z_d}^{\infty} \fr{dz}{ H(z)  \sqrt{3(R(z)+1)}}, \\
	\la{eq:DV}
	D_V(z) & \eq & \left[\fr{z}{H(z)} \left(\int_0^z \fr{dz'}{H(z')}\right)^2 \right]^{1/3}
\eea

While the sound horizon $s(z_d) \equiv s_d$ depends upon the physics prior to the recombination era, given by $z_d \approx 1059$ \ci{Ade:2015xua} and the baryon to photon ratio, $R(z) \eq \frac{3\Omega_{\gamma}}{4\Omega_b}$, it is insensitive to the dynamics of the Dark Energy, which started to act recently and was highly subdominant at that epoch. $D_V(z)$ (\ref{eq:DV}), on the other hand   is sensitive to the physics of much lower redshifts, particularly to those censed  by Large Scale Structure experiments.
 Such  galaxy surveys often measure the distance ratio $\rbao(z)$ as given by equation \ref{eq:rbao}. However, it is also common to found the inverse ratio reported or measurements of the ratios $D_A(z)/s_d$  and $D_H(z)/s_d$ where $D_A(z)$ is the angular diameter distance (\ref{eq:daz}) and $D_H(z)\eq c/H(z)$.

Many experiments (\ci{Beutler:2011hx}, \ci{Kazin:2014qga}, \ci{Anderson:2013oza}, \ci{Ross:2014qpa}, \ci{Font-Ribera:2013wce}, \ci{Delubac:2014aqe}) have measured this characteristic rule with increasing precision, providing a robust way to test the dynamics of DE through the study of the Large Scale Structure of the Universe. Upcoming experiments like DESI (Dark Energy Spectroscopic Instrument \ci{Levi:2013gra}) will probe the effects of dark energy on the expansion history using the BAO signature.

\be
	\la{eq:daz}
	D_A(z)=\frac{1}{1+z} \int_0^z \fr{dz'}{H(z')}
\ee

\begin{figure*}
    \centering
    \begin{subfigure}[b]{0.49\textwidth}
     \includegraphics[width=\textwidth]{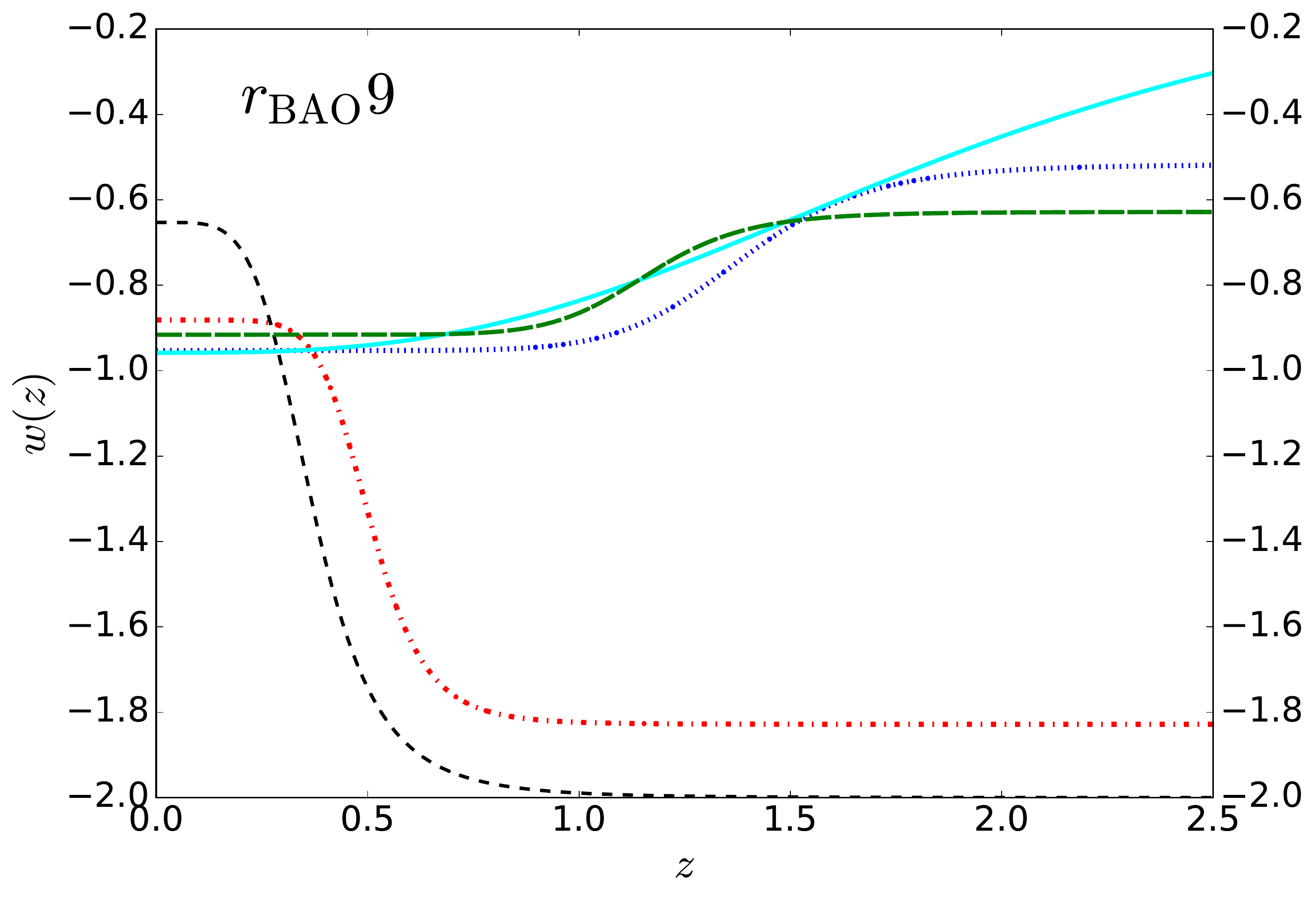}
                \end{subfigure}
      \begin{subfigure}[b]{0.49\textwidth}
        \includegraphics[width=\textwidth]{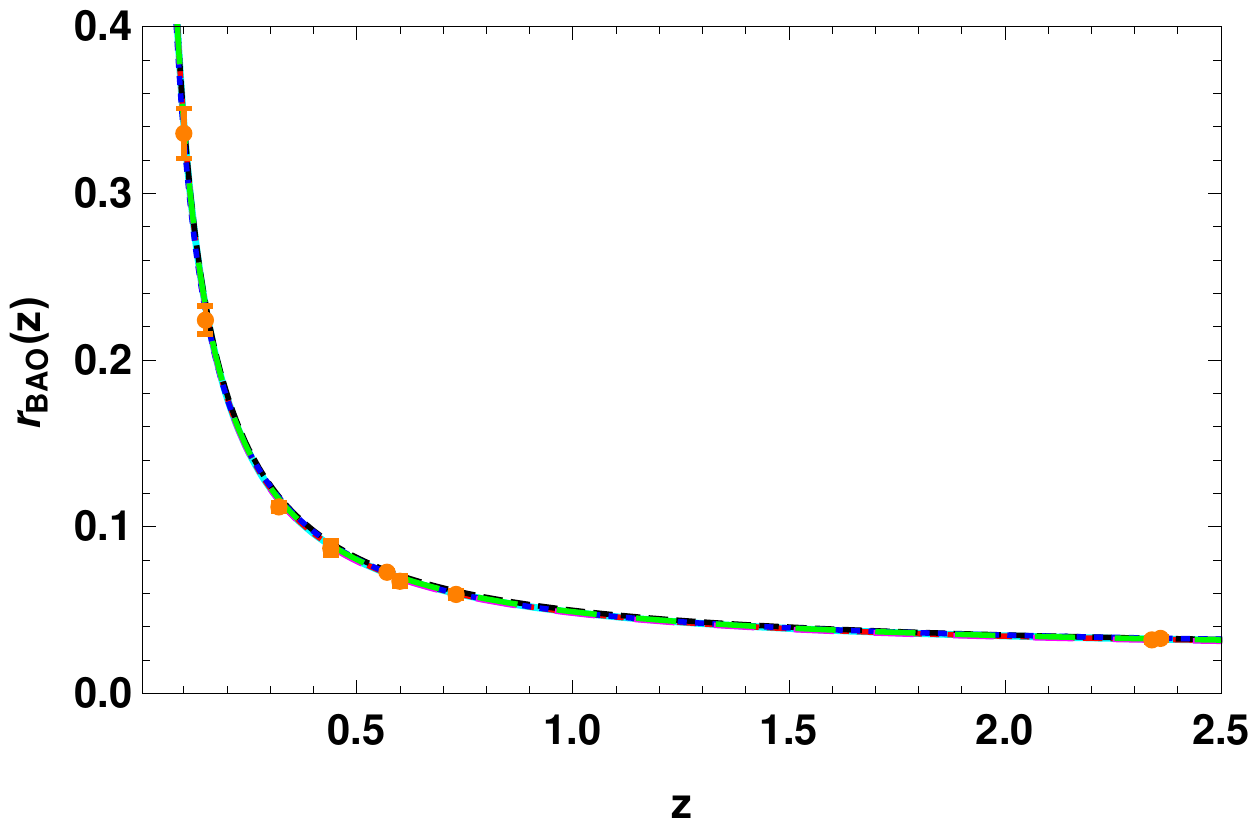}
           \end{subfigure}

    \begin{subfigure}[b]{0.49\textwidth}
        \includegraphics[width=\textwidth]{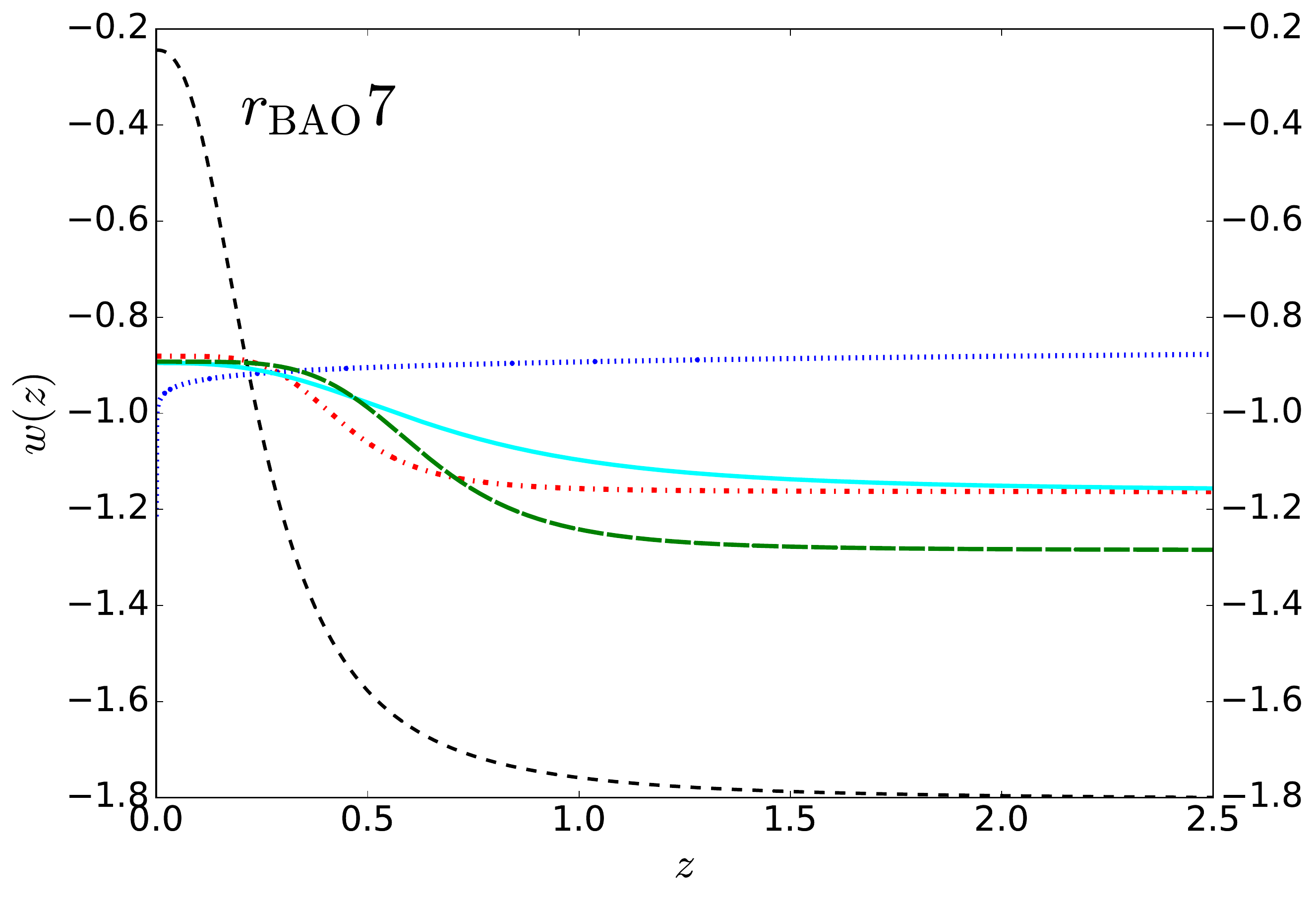}
           \end{subfigure}
      \begin{subfigure}[b]{0.49\textwidth}
       \includegraphics[width=\textwidth]{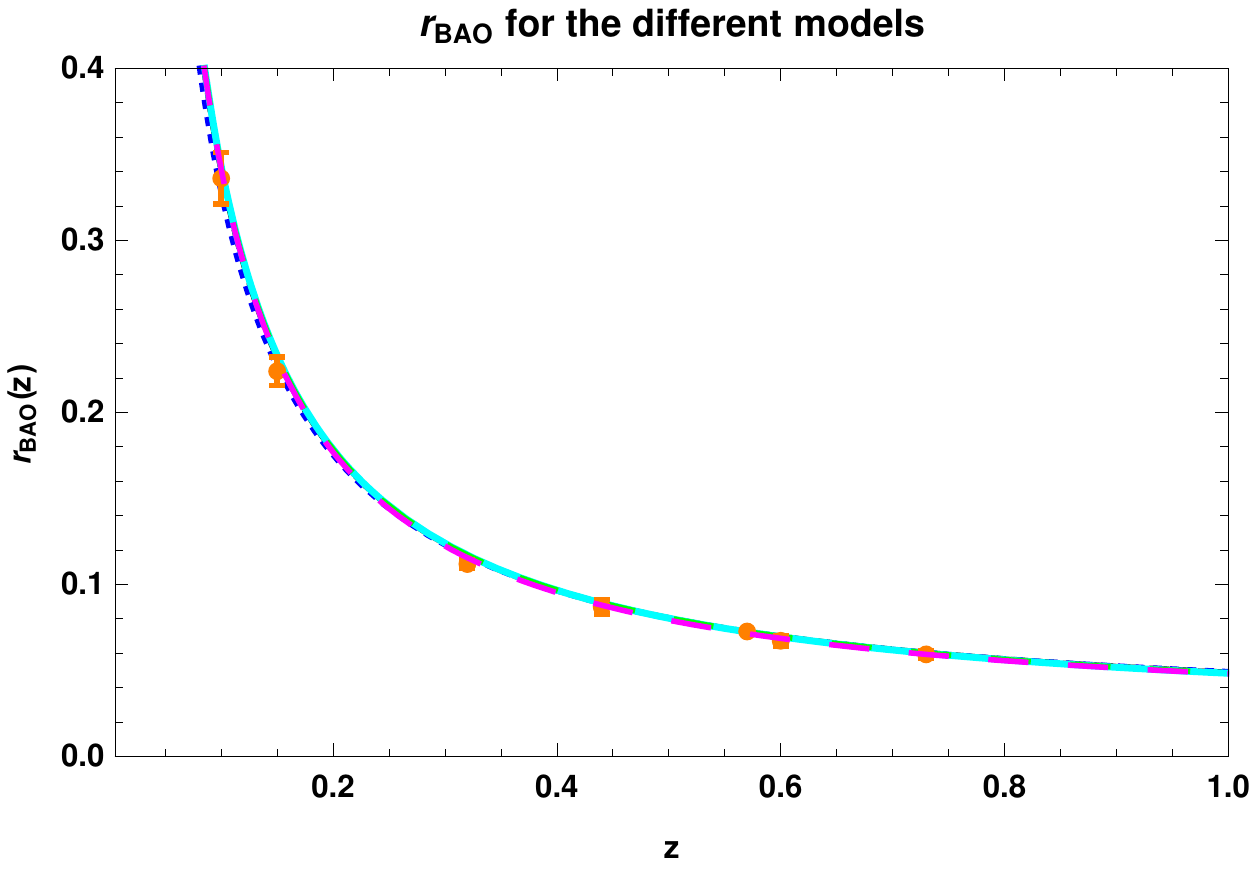}
               \end{subfigure}

    \begin{subfigure}[b]{0.49\textwidth}
        \includegraphics[width=\textwidth]{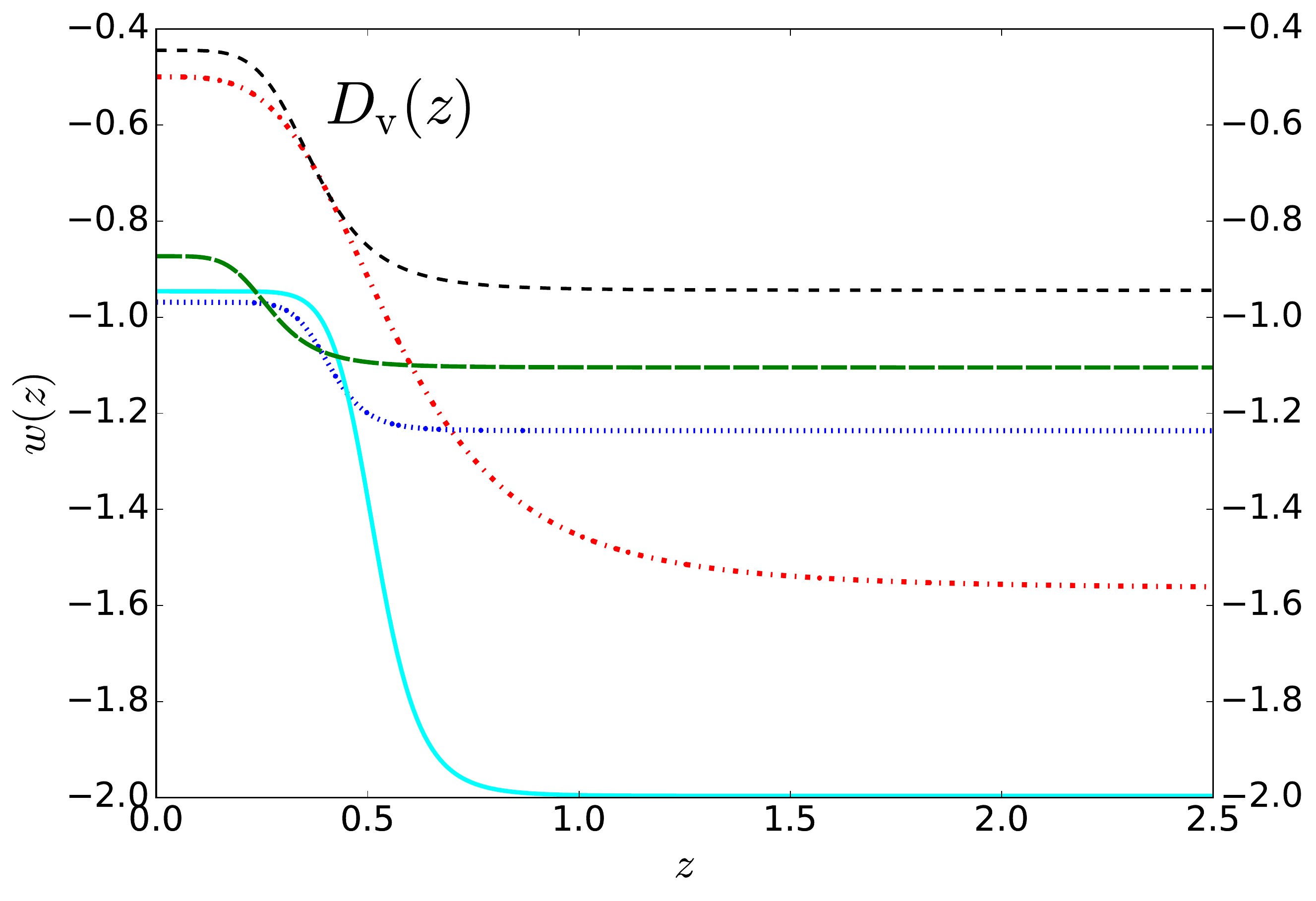}
        \caption{DE EoS behavior according to the best fit values reported in Tables \ref{table:results9pts}, \ref{table:results7pts} and \ref{table:resultsdvz}, res\-pec\-ti\-ve\-ly, from top to bottom.}
        \label{fig:bfm}
    \end{subfigure}
      \begin{subfigure}[b]{0.49\textwidth}
        \includegraphics[width=\textwidth]{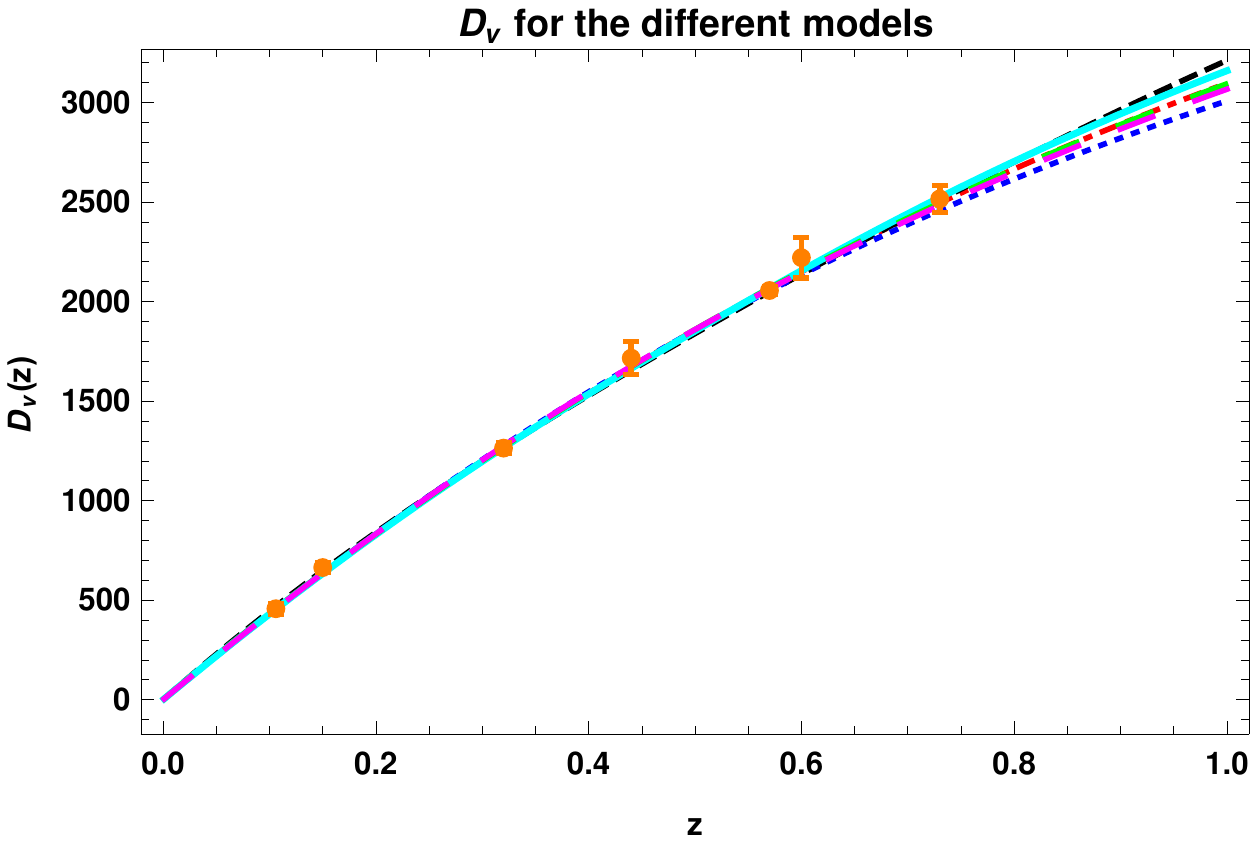}
        \caption{Plots showing the fit of the $\rbao$ and $D_V(z)$ curves corresponding to the best fit values obtained in every run to the respective set of data points. }
        \label{fig:data&curves}
    \end{subfigure}
    \caption{The curves are color coded depending on the boundaries taken during the minimization. The intervals labeled as ``$BAO_{\Om}$",  ``$BAO_{free}$", ``BAO + Planck", ``$BAO+Planck_{\Om}$ " and ``BAO + Planck $1\s$" are represented by the  dot-dashed (red), dashed (black), dotted (blue), solid (cyan) and long-dashed (green) curves, respectively. The magenta curve in the panels of column (b) represents  $\Lambda$CDM. }\label{fig:results}
\end{figure*}

\section{Analysis}
\la{sec:Analysis}

In agreement with the considerations detailed above this analysis  will use the available BAO distance data which lies in the range $z\in[0,2.36]$ combined with the priors given by the CMB as reported by  the Planck Collaboration (  P.~A.~R.~Ade {\it et al.} 2015, \ci{Ade:2015xua}).

In order to work with the reported measurements of the BAO signature we used the $\chi^2$ function defined in terms of some observational data points, $D_i$, the related theoretical prediction, $y(x_i|\vec{\theta})$ and the associated observational errors, $\s_i$.

\be
	\la{eq:chi2}
	\chi^2 \eq\sum_i\left[D_i - y(x_i|\vec{\theta})\right]^2 \left( \fr{1}{\s_i} \right)^2
\ee

By minimizing  this function, the best fit values for the parameters, $\vec{\theta}$, in the theoretical model are found.  Associated to the value of the $\chi^2$ function it is possible to define confidence regions in the parameter space to represent the standard $1\s$, $2\s$ contours.

\subsection{Observational data and numerical minimization}

We make use of the observational points from the six-degree-field galaxy survey (6dFGS
\ci{Beutler:2011hx}), Sloan Digital Sky Survey Data Release 7 and 11 (SDSS DR7 \ci{Ross:2014qpa} and SDSS DR11 \ci{Anderson:2013oza}), the WiggleZ dark energy survey (\ci{Kazin:2014qga}) and the Lymann $\a$ Forest (Ly$\a$-F) measurements from the Baryon Oscillation Spectroscopic Data Release 11 (BOSS DR11 \ci{Font-Ribera:2013wce}, \ci{Delubac:2014aqe}) as analyzed and reported by Gong {\it et al} 2015 \ci{Gong:2015tta}. Table \ref{table:datarbao} summarizes them all.  The associated $D_V(z)$ values are included in Table \ref{table:datadvz}.

Additionally to the free parameters in the  equation \ref{eq:eos} we also investigated the cons\-traints on  $\Om$  and $H_0$ (or equivalently $h$),  resulting in  the set   $\vec{\theta}$ = $\{w_0,  w_i,  z_t,  q,  \Om,  h \}$.

The latest results from the Planck Collaboration (  P.~A.~R.~Ade {\it et al.} 2015, \ci{Ade:2015xua}) were used to fix the fiducial cosmological parameters. In particular, we chose the results when the full CMB polarization was used (without adding lensing data). This is, we are using the Planck (TT + TE + EE + low P) values from the table 4 of \ci{Ade:2015xua}.

The minimization of the $\chi^2$ function was done within the boundaries indicated in Table \ref{table:boundaries} and using the observational data from Tables \ref{table:datarbao} and \ref{table:datadvz}. The software Mathematica 10.1 was used to perform the constrained minimization numerically.

The fiducial values for $\Om$ and $h$ are denoted by $\overline{\Omega}_m = 0.3156$ and $\overline{h}=0.6727$, respectively, corresponding to the central values as reported by \ci{Ade:2015xua}.

We considered five di\-ffe\-rent priors; for each one of the three different data sets ($r_{BAO}(z)$  including Ly$\a$-F measurements, $r_{BAO}(z)$ without Ly$\a$-F measurements and $D_V(z)$ data) we varied the free parameters of the EoS within the limits summarized in Table \ref{table:boundaries} using different intervals for $\Omega_m$ and $h$:

\begin{enumerate}
\item ``BAO+Planck" corresponds to fix $\Omega_m = \overline{\Omega}_m$ and  $h=\overline{h}$ to the central values from Planck.
\item ``BAO + Planck$_{\Omega_m}$" corresponds to fix $h=\overline{h}$ but let $\Omega_m$ to vary within $\pm 1\sigma$ limits from Planck.
\item ``BAO + Planck1$\s$" was for the case both $\Omega_m$ and $h$ varying within their $\pm 1\s$  Planck error intervals.
\item ``BAO$_{\Omega_m}$" corresponds to fix $h=\overline{h}$ and let $\Omega_m$ to vary between [0, 1] 
\item ``BAO$_{free}$" stands for the case when $\Omega_m$ was left to vary in [0, 1] and $h$ in [0.5, 1].
\end{enumerate}

\begin{table}
	\begin{center}
		\begin{tabular}{ |c|c|c|}
		\hline
\bf{Data set} & \bf{Redshift} &$\rbao(z)$ \\ \hline
6dF  & 0.106 \ci{Beutler:2011hx} & 0.336 $\pm$ 0.015 \\ \hline
SDSS DR7  & 0.15 \ci{Ross:2014qpa} &  0.2239 $\pm$ 0.0084 \\ \hline
\multirow{3}{*}{WiggleZ} & 0.44  \ci{Kazin:2014qga}& 0.0870 $\pm$ 0.0042 \\ & 0.60 \ci{Kazin:2014qga} & 0.0672 $\pm$ 0.0031  \\ & 0.73  \ci{Kazin:2014qga}& 0.0593 $\pm$ 0.0020  \\ \hline
\multirow{4}{*}{SDSS-III DR11} & 0.32 \ci{Anderson:2013oza} & 0.1181 $\pm$ 0.0023   \\ & 0.57 \ci{Anderson:2013oza}& 0.0726 $\pm$ 0.0007  \\ & 2.34 \ci{Delubac:2014aqe} & 0.0320 $\pm$ 0.0013 \\ & 2.36  \ci{Font-Ribera:2013wce} & 0.0329 $\pm$ 0.0009 \\ \hline
		\end{tabular}
		\caption{$\rbao(z)$ measurements as reported by \ci{Gong:2015tta}: the values for SDSS data were inverted from the published values of $D_V(Z)/s_d$ and the values corresponding to the Ly$\a$-F data were obtained from the reported quantities $D_A(z)/s_d$ and $D_H(z)/s_d$. }
		\la{table:datarbao}
	\end{center}
\end{table}

\begin{table}
	\begin{center}
		\begin{tabular}{ |c|c|c|}
		\hline
\bf{Data set} & \bf{Redshift}  & $D_V(z) (Mpc)$\\ \hline
6dF  & 0.1 \ci{Beutler:2011hx}  & 457 $\pm$ 27\\ \hline
SDSS DR7  & 0.15 \ci{Ross:2014qpa}  & 664 $\pm$ 25\\ \hline
\multirow{3}{*}{WiggleZ} & 0.44  \ci{Kazin:2014qga}&  1716 $\pm$ 83\\ & 0.60 \ci{Kazin:2014qga}  & 2221 $\pm$ 101 \\ & 0.73  \ci{Kazin:2014qga}& 2516 $\pm$ 68 \\ \hline
\multirow{2}{*}{SDSS-III DR11} & 0.32 \ci{Anderson:2013oza}  & 1264 $\pm$ 25 \\ & 0.57 \ci{Anderson:2013oza}&   2056  $\pm$ 20 \\ \hline
		\end{tabular}
		\caption{$D_V(z)$ measurements as reported by every experiment. }
		\la{table:datadvz}
	\end{center}
\end{table}

\begin{table}
	\begin{center}
		\begin{tabular}{|c|c|c|}
		\hline
		\bf{Parameter} & \bf{Interval 1} & \bf{Interval 2}\\
		\hline
		$w_0$ & $[-2,0]$ & $[-2,0]$   \\
		$w_i$ & $[-2,0]$  & $[-2,0]$  \\
		$z_T$ & $[0.01,3]$ & $[0.01,3]$ \\
		$q$ & $[0.1,10]$ & $[0.1,10] $ \\
		$\Om$ & $[0,1]$ & $[0.3065, 0.3247]$ \\
		$h$	& $[0.5,1]$ & $[0.6661, 0.6793]$\\ \hline
		\end{tabular}
		\caption{Boundaries for the parameters constrained. The second and third column differ from one another by the boundaries imposed to $\Om$ and $h$. The later uses the $\pm 1\s$ values corresponding to the full CMB data  (TT, TE, EE + lowP without lensing)  as reported by the Planck collaboration (Table 4 of \ci{Ade:2015xua}). }
		\la{table:boundaries}
	\end{center}
\end{table}

In order to asses the goodness of the performed fit we  use of the so called reduced chi-square function, $\chi^2_{red}\eq \chi^2/\nu$, where $\nu$ represents the number of degrees of freedom, defined to be the difference between the number of observational points, N,  and the parameters fitted, M, \emph{i. e.},  $\nu \eq N- M$.  The results obtained for each set of runs are displayed  in Tables \ref{table:results9pts}-\ref{table:resultsdvz} where the best fit for the constrained parameters and the value for the chi-squared function are reported.

Tables \ref{table:results9pts}-\ref{table:resultsdvz}  also include the corresponding $\chi^2$  value for $\La$CDM, when $\Om$ and $h$ were kept fixed to their fiducial values and the parameters of the EoS were reported in the form $\{w_0=-1, w_i=-1, q=1, z_t=1\}$ to represent $w=-1$.

\begin{figure*}
	\centering
		 \begin{subfigure}[t]{0.45\textwidth}	
		 \captionsetup{width=5cm}
	        \includegraphics[width=\textwidth]{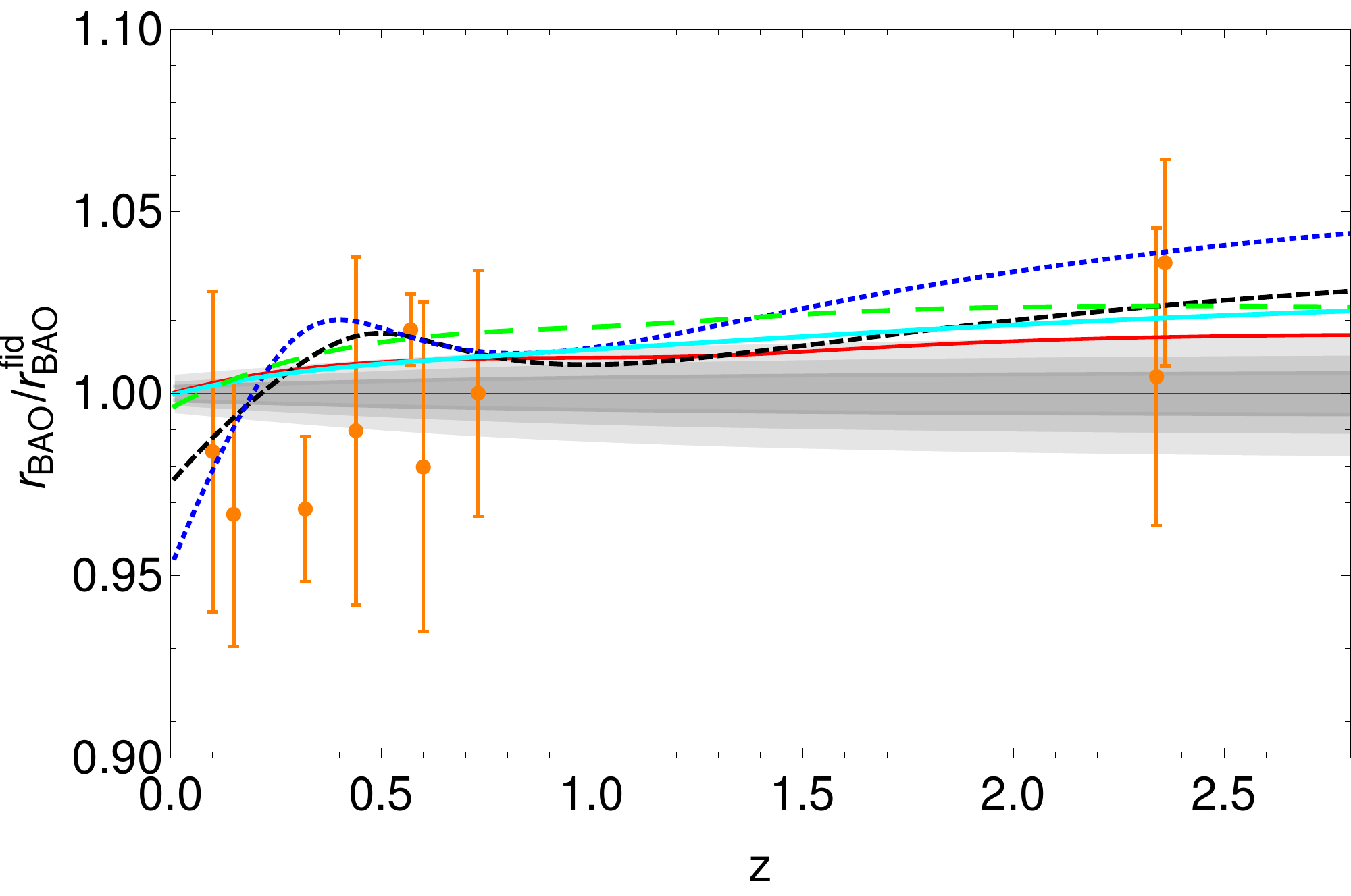}
        		\label{fig:ratio9}
        		\caption{}
    \end{subfigure}\quad
     \begin{subfigure}[t]{0.45\textwidth}
	 \captionsetup{width=5cm}
        \includegraphics[width=\textwidth]{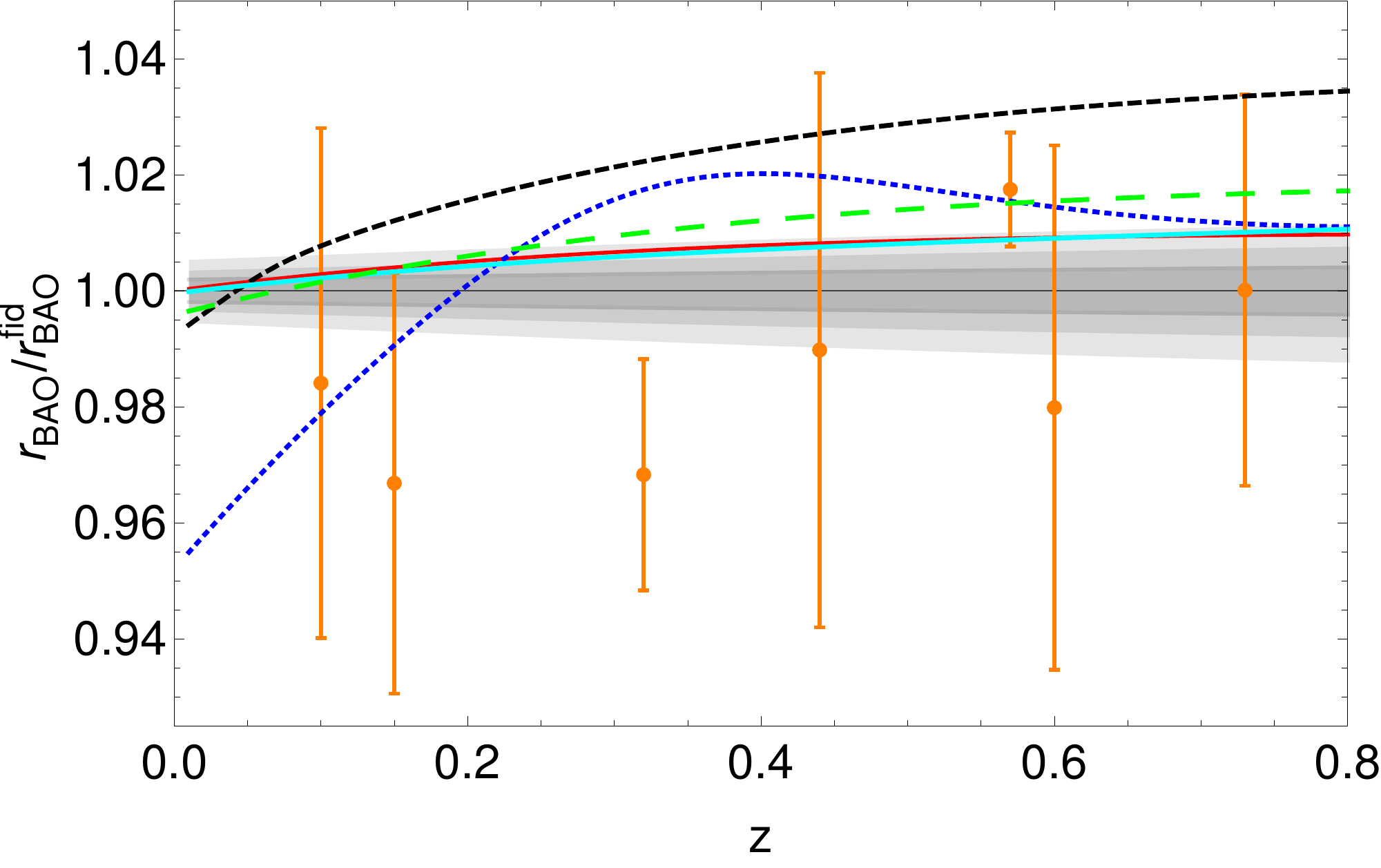}
        \label{fig:ratio7}
        \caption{}
    \end{subfigure}\quad

    \centering
		 \begin{subfigure}[t]{0.45\textwidth}
	 \captionsetup{width=5cm}
        \includegraphics[width=\textwidth]{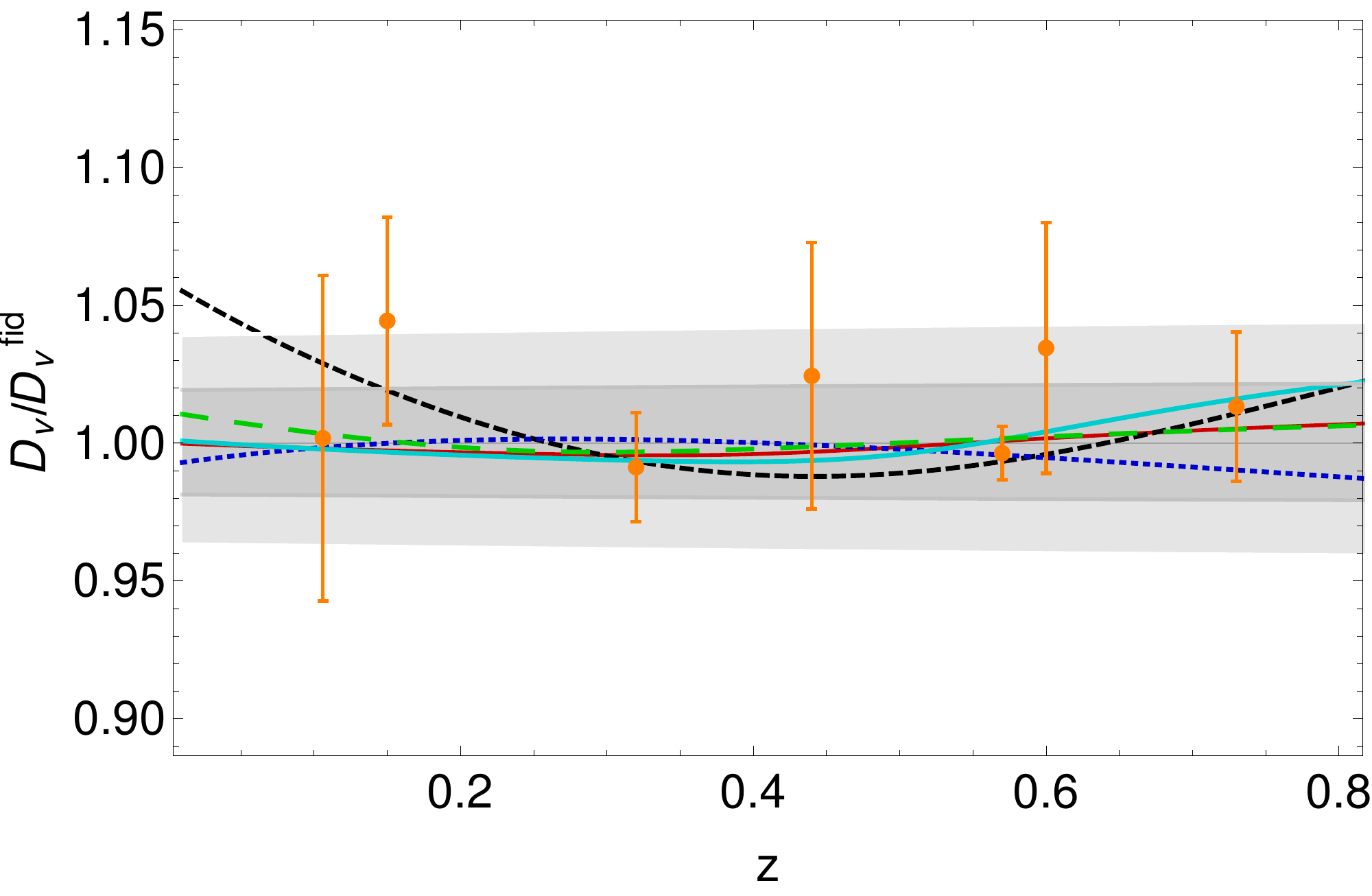}
        \label{fig:ratioDVz}
        \caption{}
    \end{subfigure}\quad

\caption{Comparison of both the best fit models to the fiducial one and the measurements  reported by the experiments when assuming $\La$CDM. The curves are color coded depending on the boundaries taken during the minimization. The intervals labeled as ``$BAO_{\Om}$",  ``$BAO_{free}$", ``BAO + Planck", ``$BAO+Planck_{\Om}$" and ``BAO + Planck $1\s$" are represented by the  red, black, blue, cyan and green curves, respectively, and the gray bands represent  the 1$\s$, 2$\s$, 3$\s$ regions estimated as explained in the text.}
	\label{fig:ratios}
\end{figure*}

\begin{figure*}
	\centering
	
	 \begin{subfigure}[t]{0.3\textwidth}
	 \captionsetup{width=4cm}
        \includegraphics[width=0.95\textwidth]{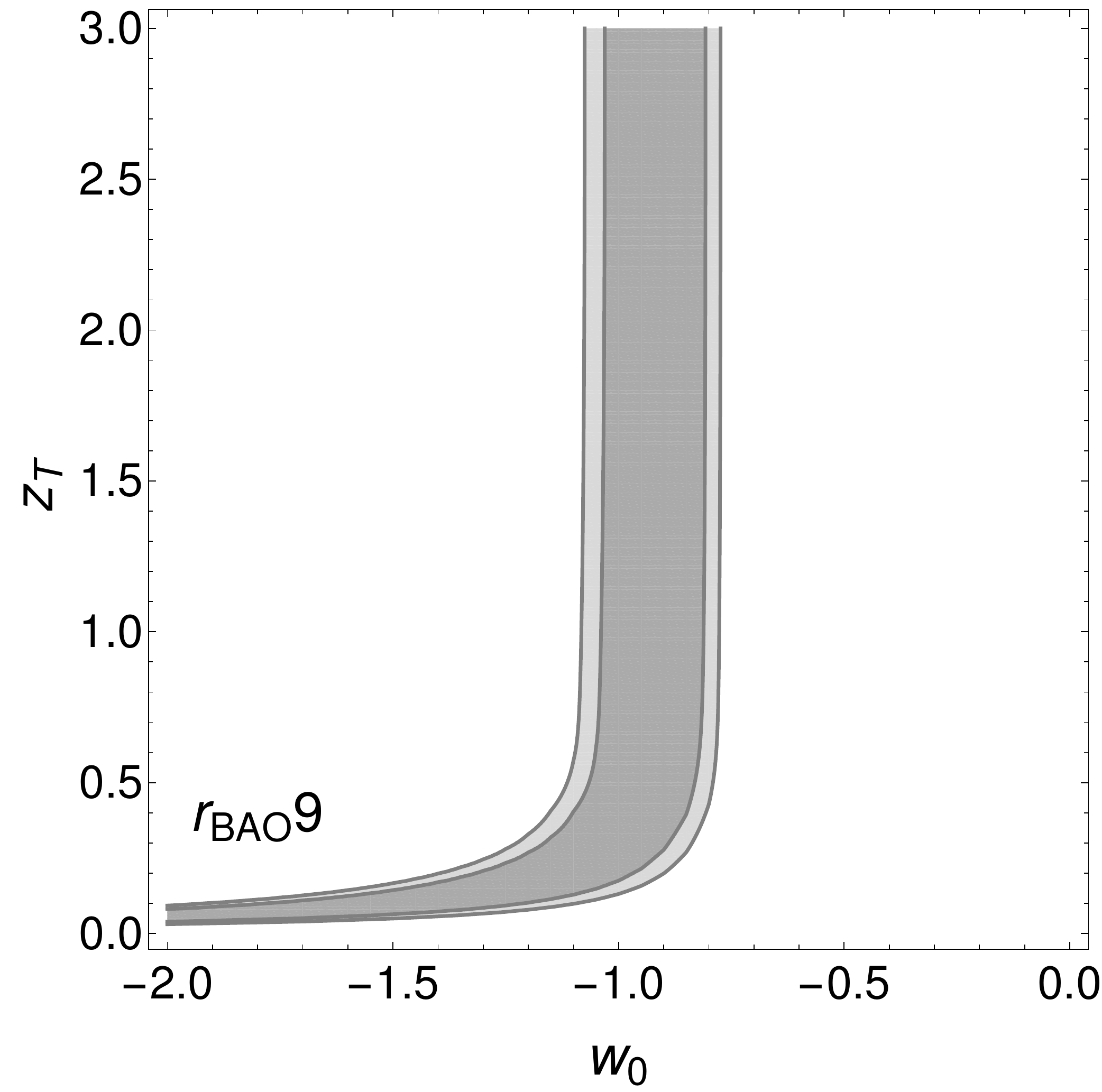}
        \label{fig:w0-zT-9}
        \caption{}
    \end{subfigure}\quad
    \begin{subfigure}[t]{0.32\textwidth}
	\includegraphics[width=0.9\linewidth]{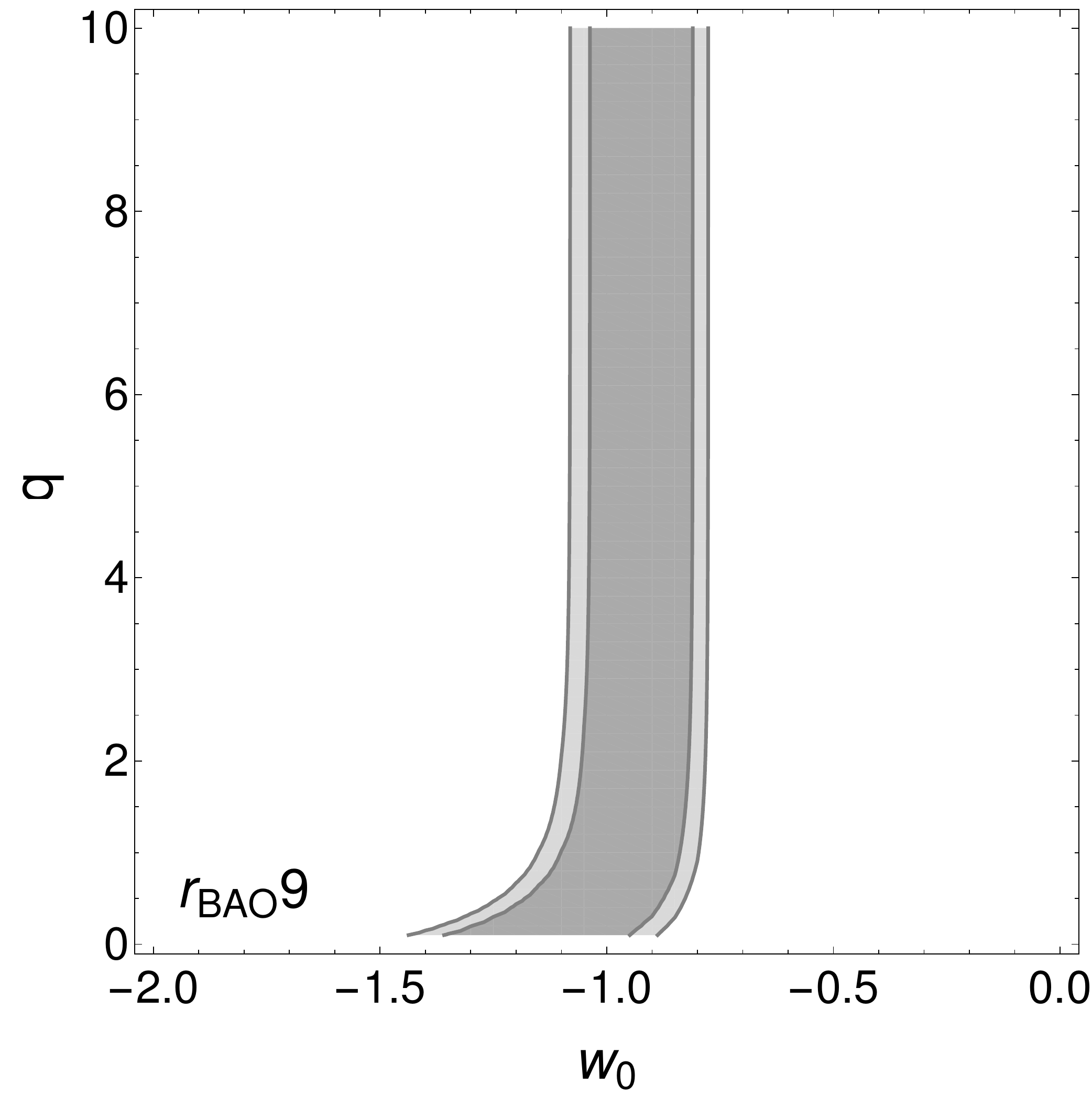}
\label{fig:w0-q-9}
        \caption{}
    \end{subfigure}\quad
        \begin{subfigure}[t]{0.32\textwidth}
	\includegraphics[width=0.9\linewidth]{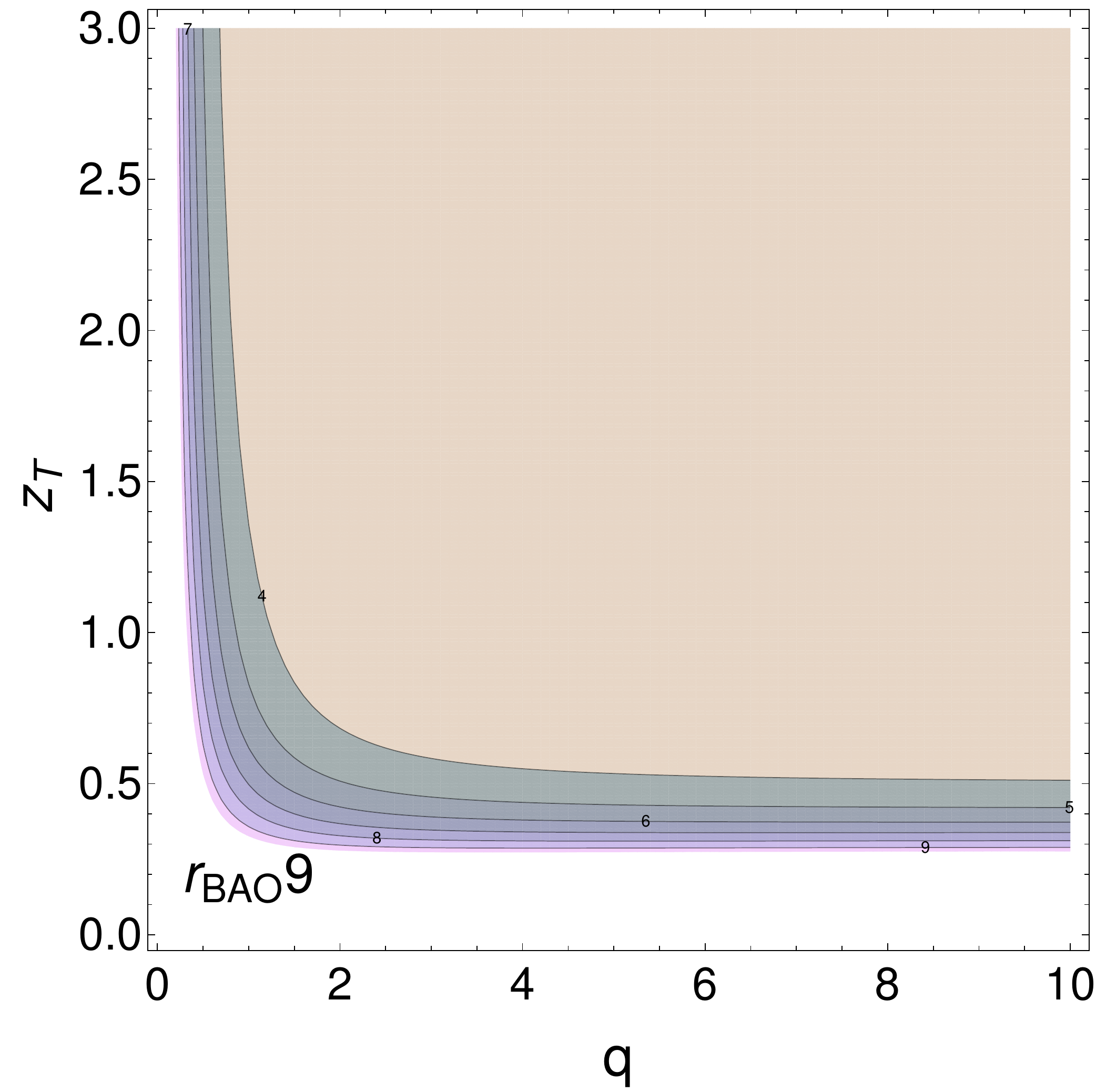}
\label{fig:q-zT-9}
        \caption{}
    \end{subfigure}\quad

 \begin{subfigure}[t]{0.3\textwidth}
	 \captionsetup{width=4cm}
        \includegraphics[width=0.95\textwidth]{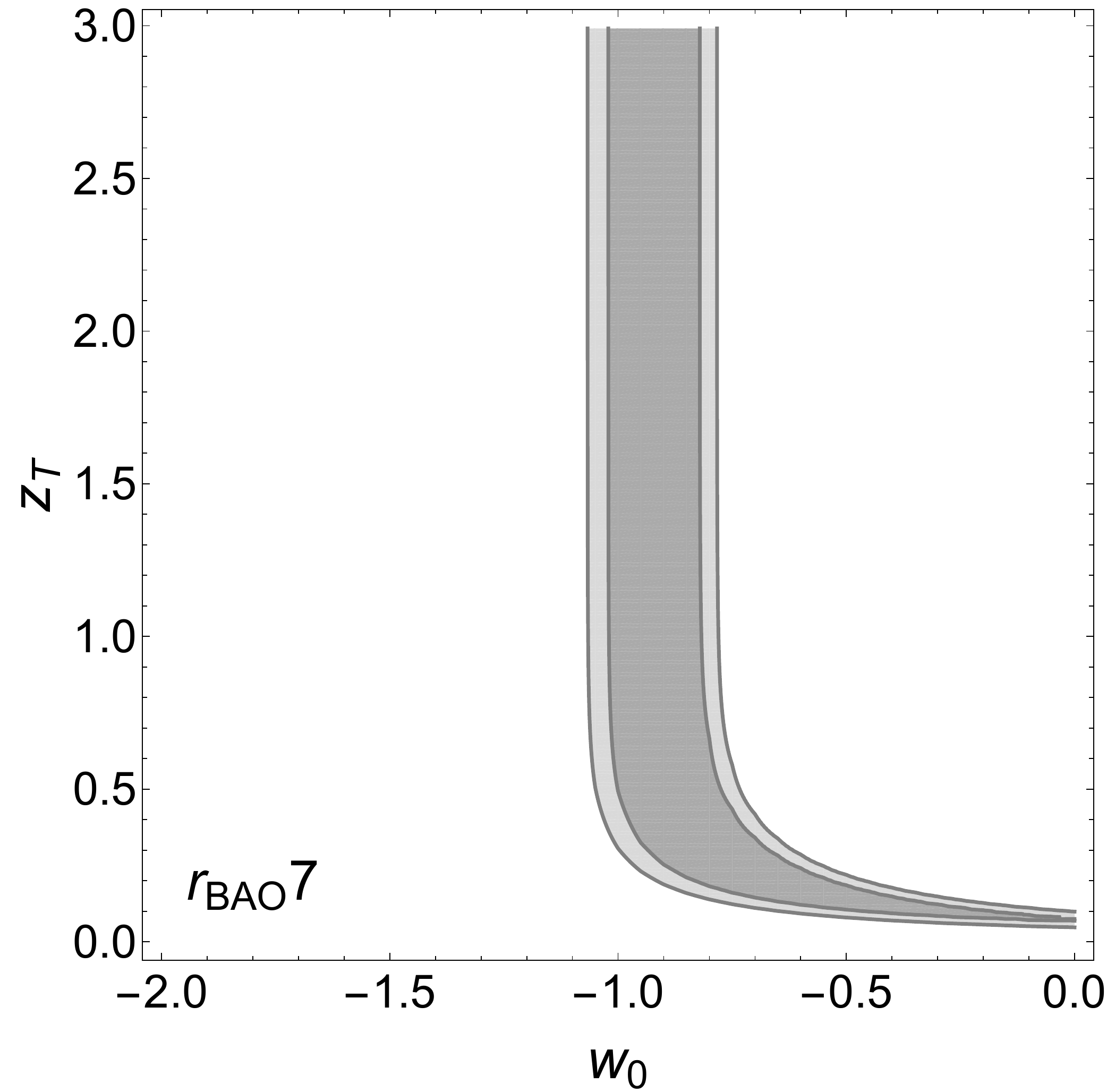}
        \label{fig:w0-zT-7}
        \caption{}
    \end{subfigure}\quad
    \begin{subfigure}[t]{0.32\textwidth}
	\includegraphics[width=0.9\linewidth]{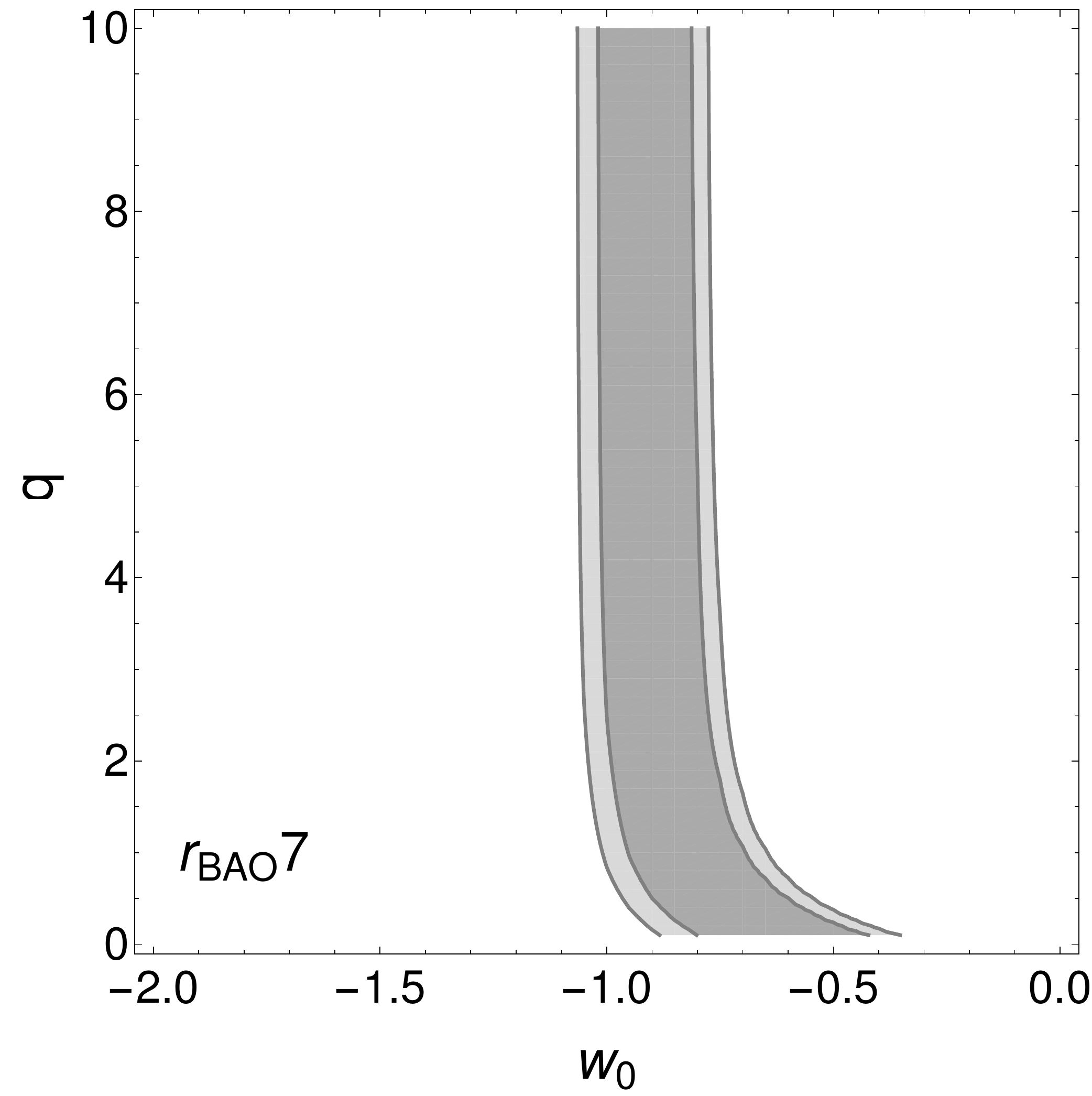}
\label{fig:w0-q-7}
        \caption{}
    \end{subfigure}\quad
        \begin{subfigure}[t]{0.32\textwidth}
	\includegraphics[width=0.9\linewidth]{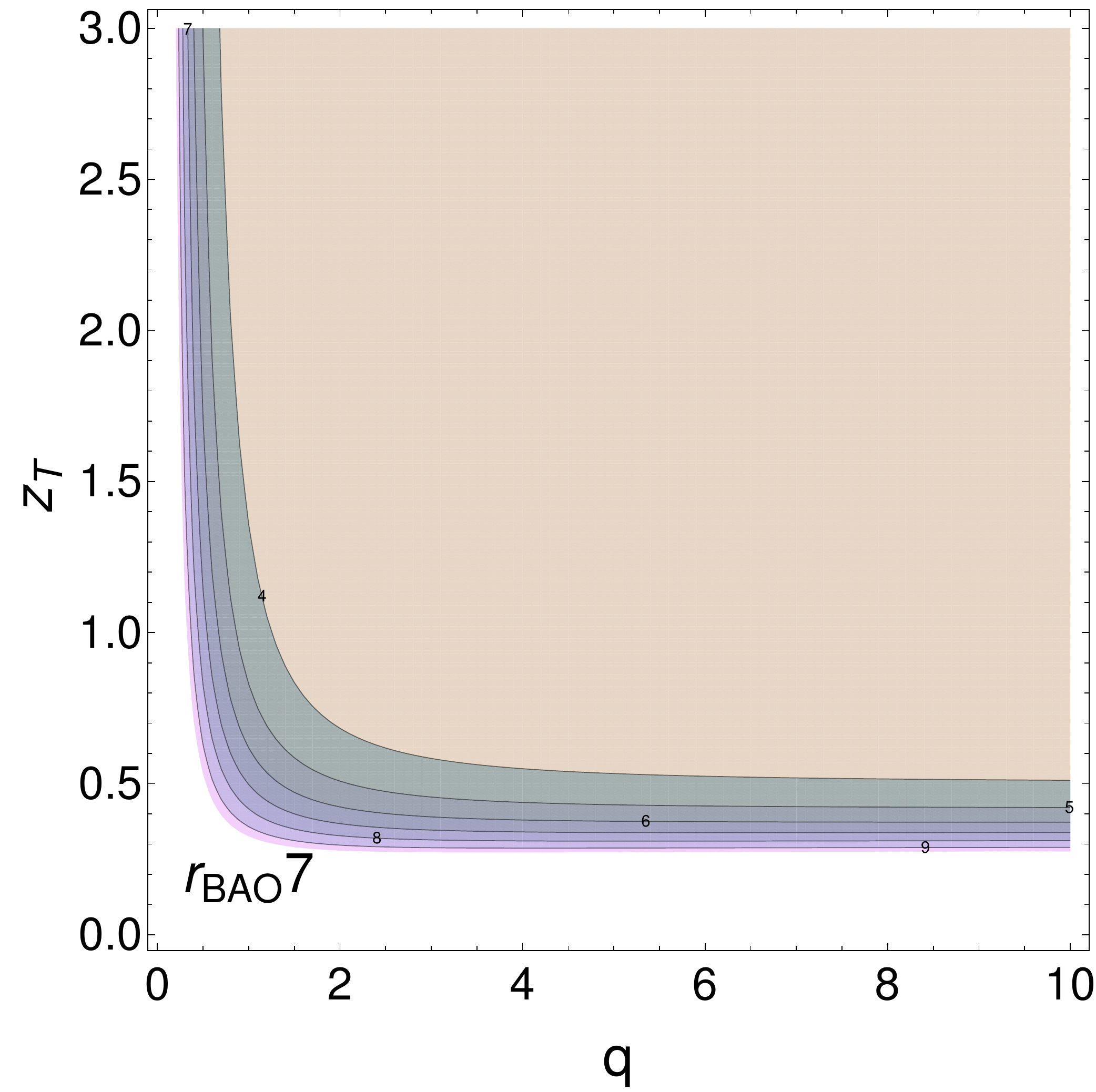}
\label{fig:q-zT-7}
        \caption{}
    \end{subfigure}\quad
    
\begin{subfigure}[t]{0.3\textwidth}
	 \captionsetup{width=4cm}
        \includegraphics[width=0.95\textwidth]{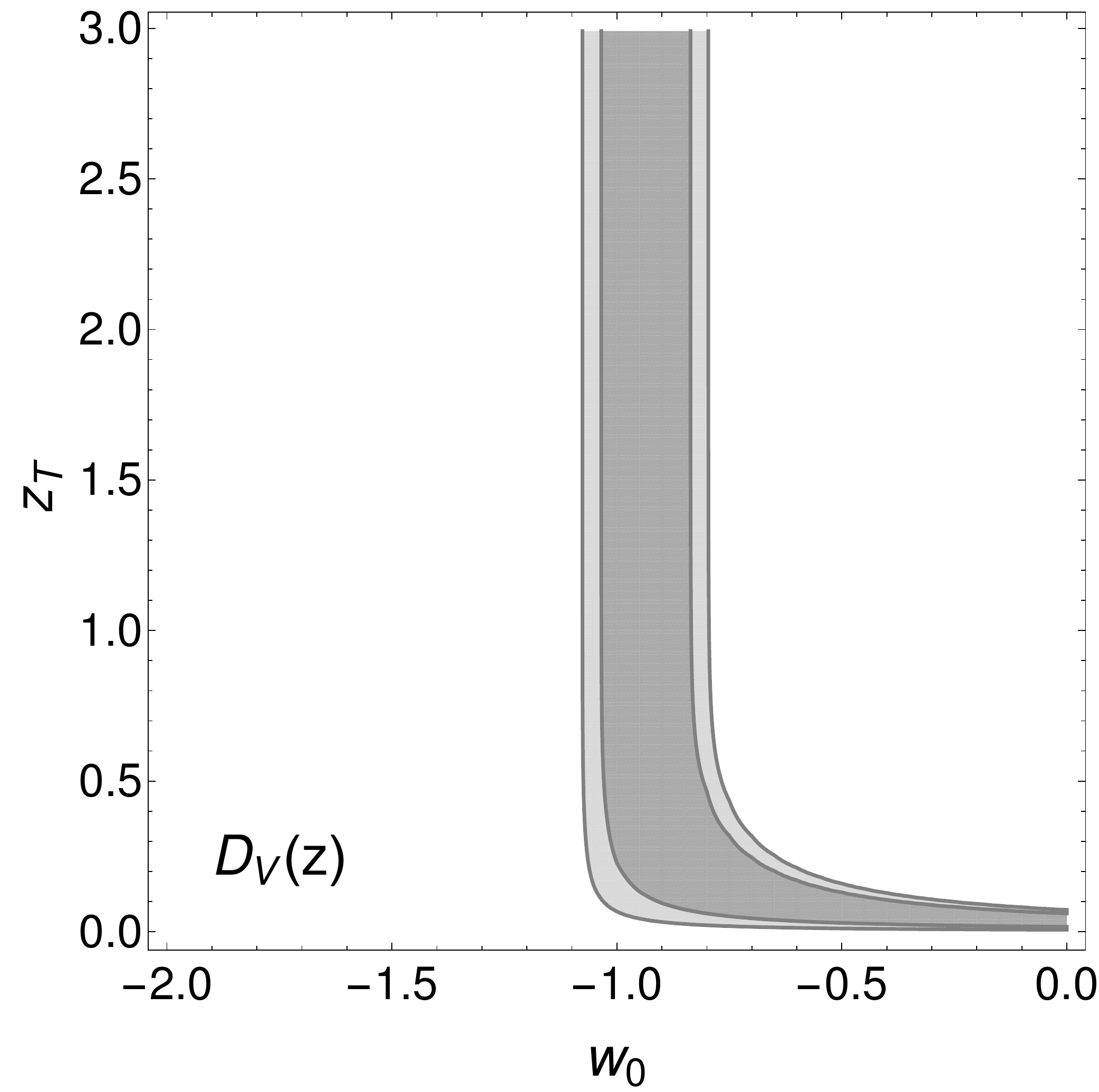}
        \label{fig:w0-zT-DvZ}
        \caption{}
    \end{subfigure}\quad
    \begin{subfigure}[t]{0.32\textwidth}
	\includegraphics[width=0.9\linewidth]{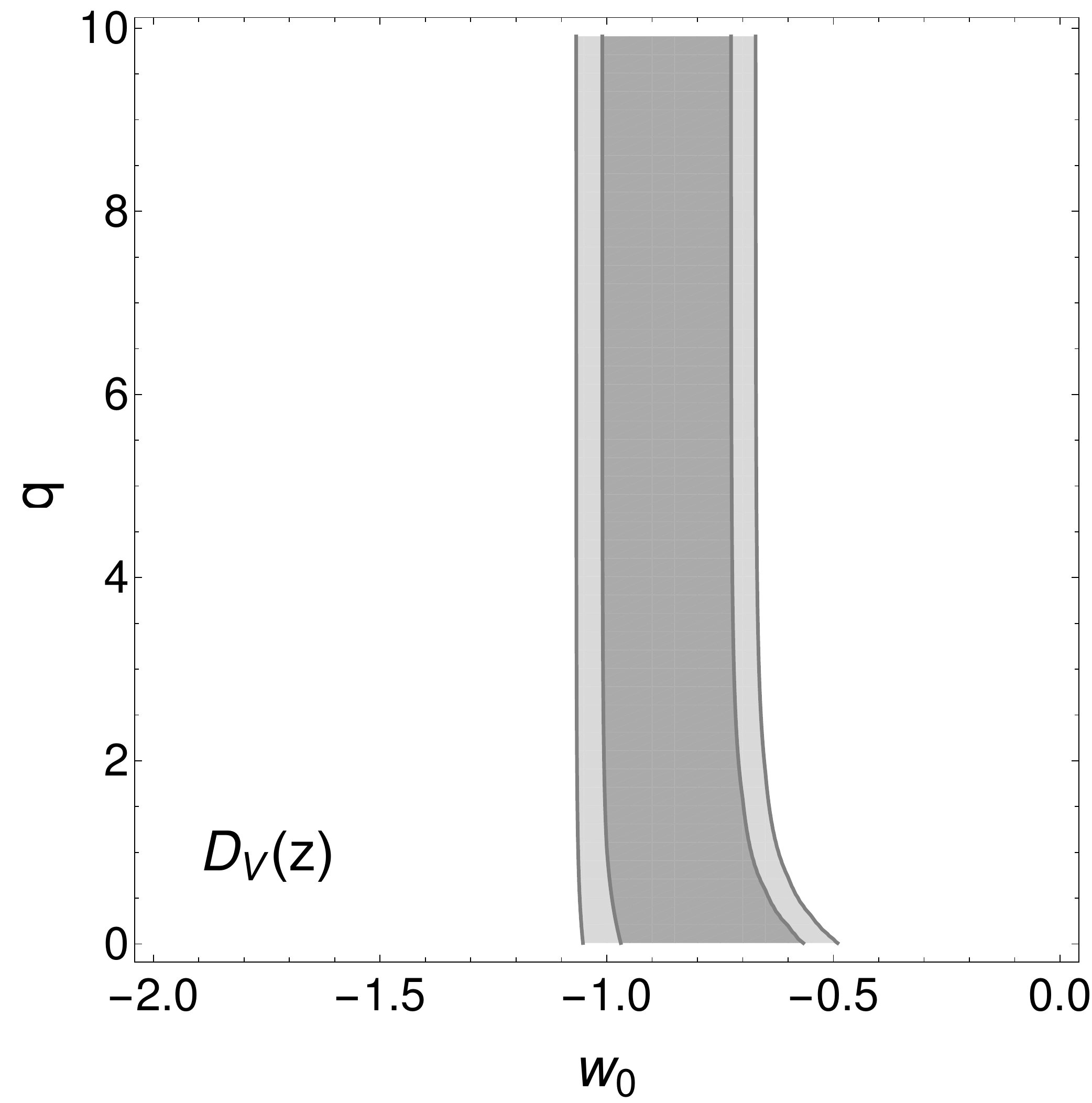}
\label{fig:w0-q-DvZ}
        \caption{}
    \end{subfigure}\quad
        \begin{subfigure}[t]{0.32\textwidth}
	\includegraphics[width=0.9\linewidth]{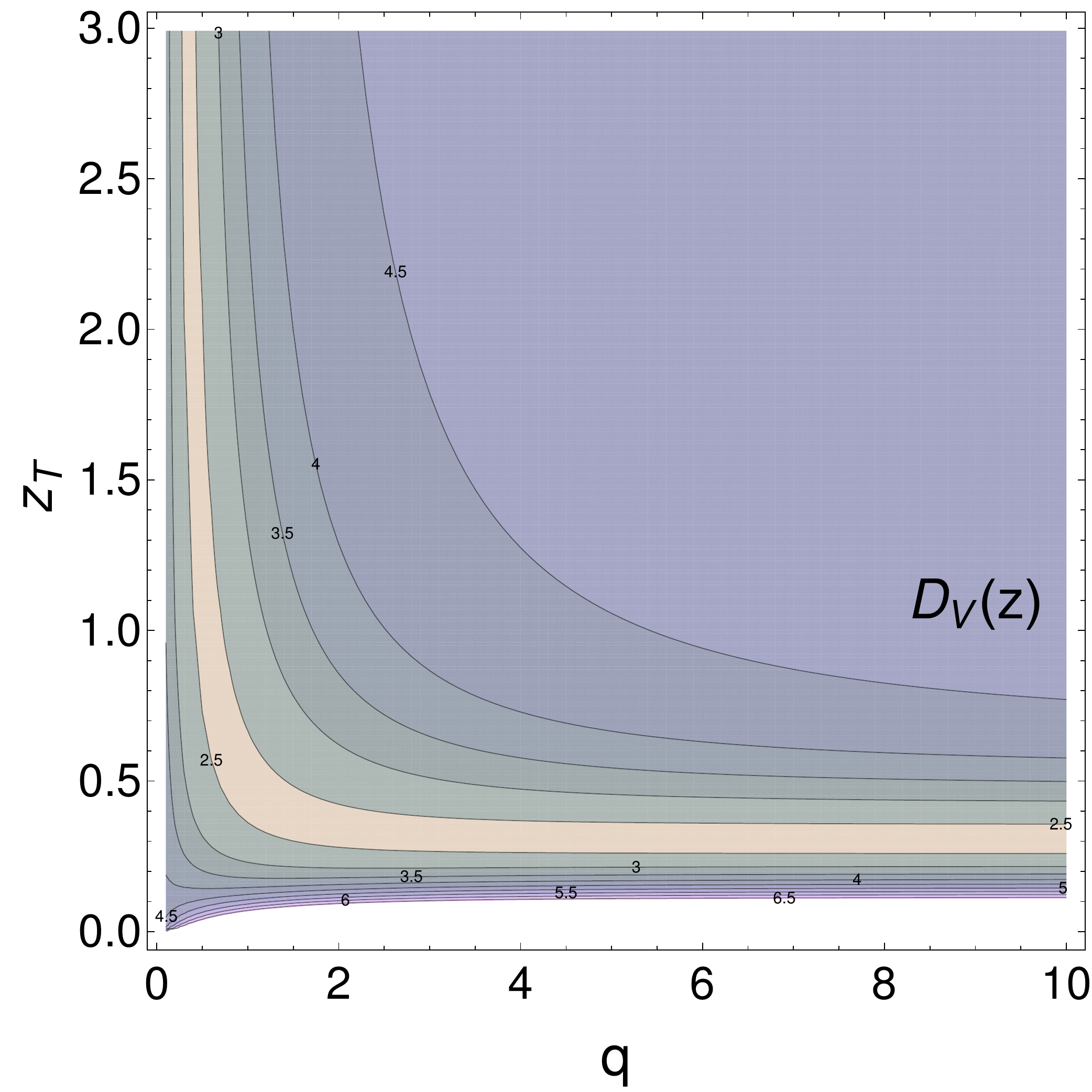}
\label{fig:q-zT-DvZ}
        \caption{}
    \end{subfigure}\quad

	\caption{Contour plots for the ``BAO+Planck 1$\sigma$" runs displaying the 1$\sigma$-2$\sigma$ confidence levels for the parameters $w_0$-$q$ and $w_0$-$z_T$.  The $\chi^2$ contours around the minimum value for the $q$-$z_T$ space are also shown.}
	\label{fig:w0-zT-q}
\end{figure*}

\begin{figure*}[h!]
	\centering
		 \begin{subfigure}[t]{0.35\textwidth}
	 \captionsetup{width=5cm}
        \includegraphics[width=\textwidth]{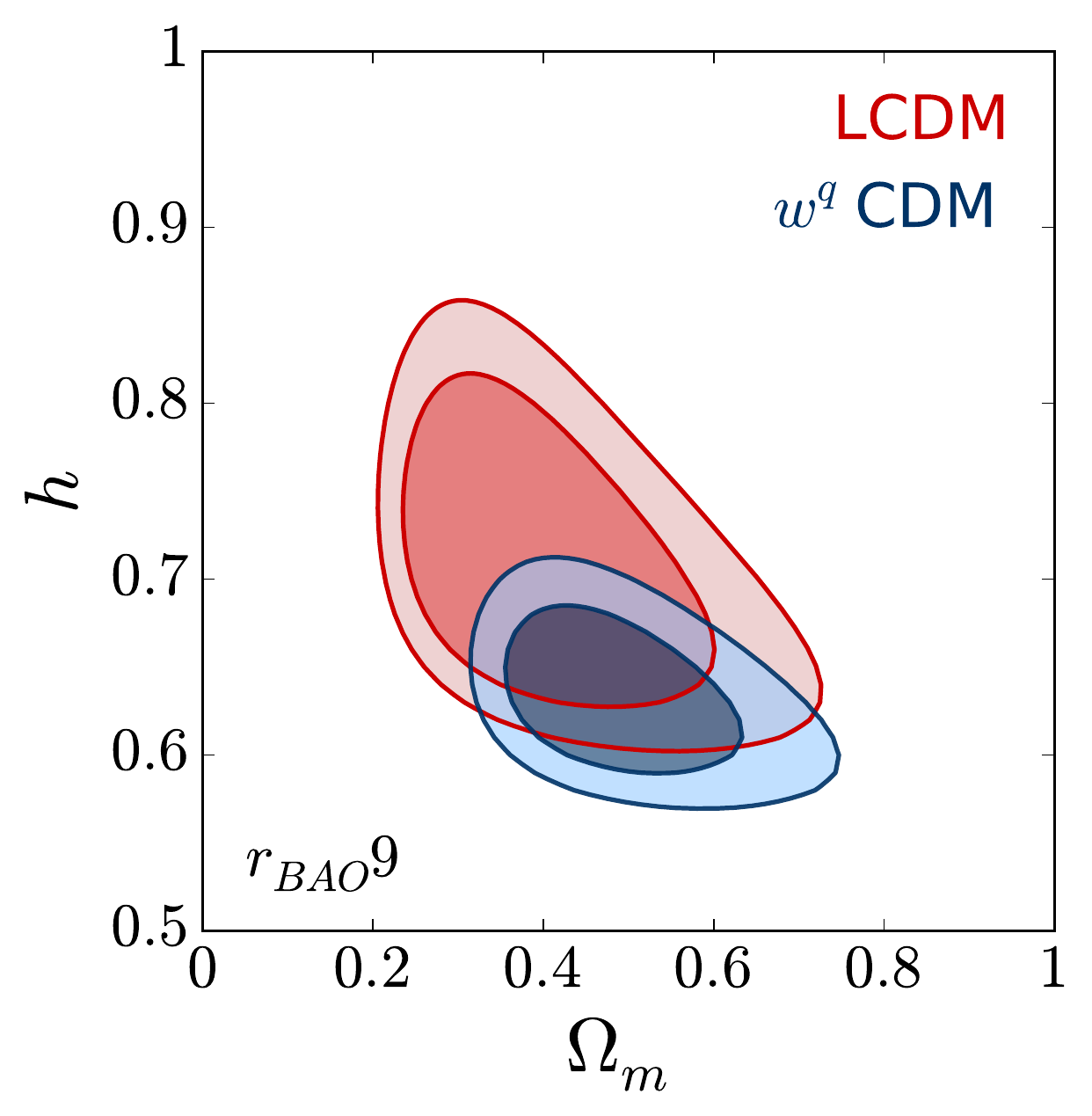}
        \label{fig:BAO3-LCDMFree_9}
        \caption{}
    \end{subfigure}\quad
     \begin{subfigure}[t]{0.39\textwidth}
	\includegraphics[width=0.9\linewidth]{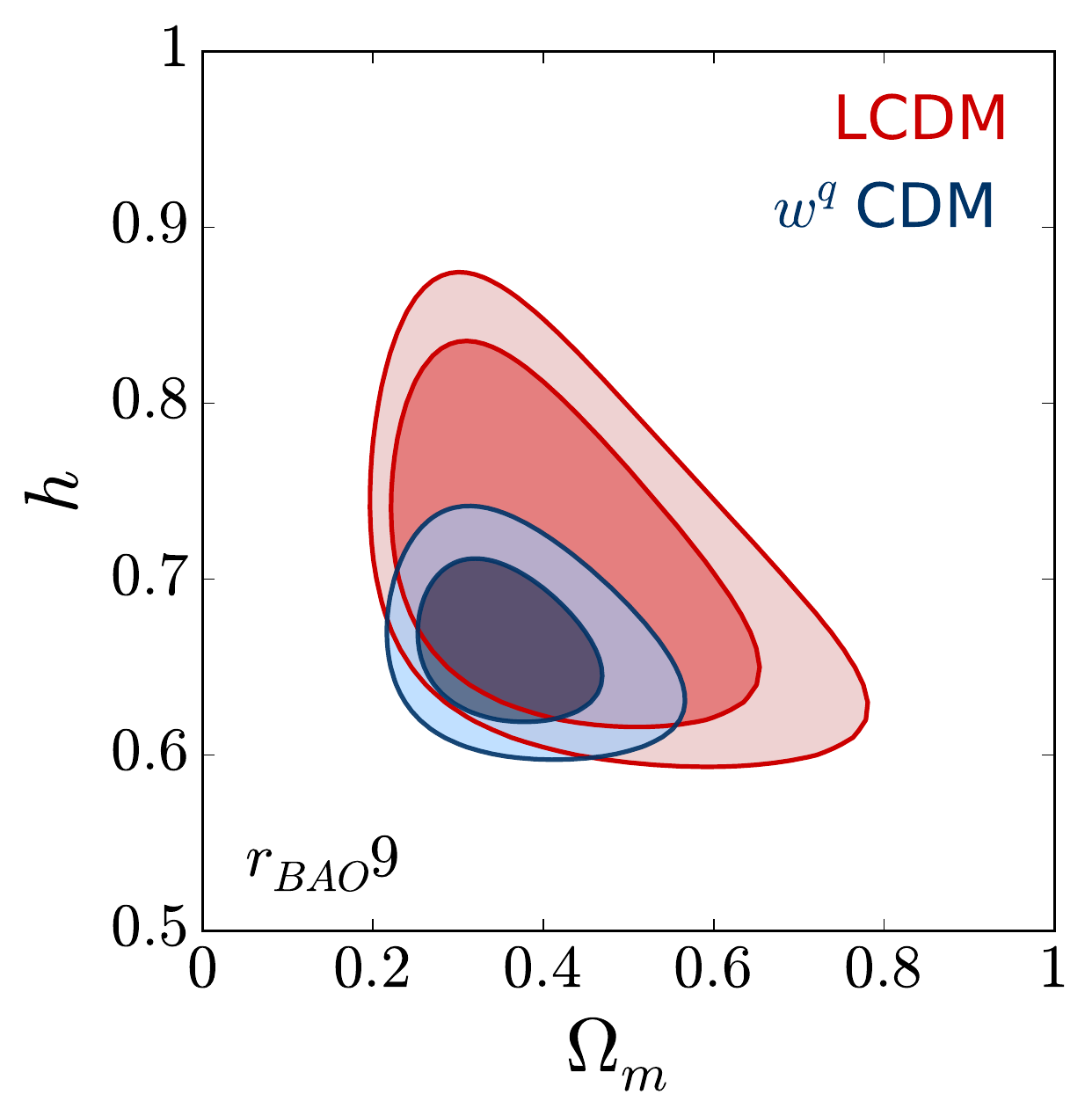}
\label{fig:BAOPlanck2-LCDMPlanck_9}
        \caption{}
    \end{subfigure}\quad

     \begin{subfigure}[t]{0.35\textwidth}
	 \captionsetup{width=5cm}
        \includegraphics[width=\textwidth]{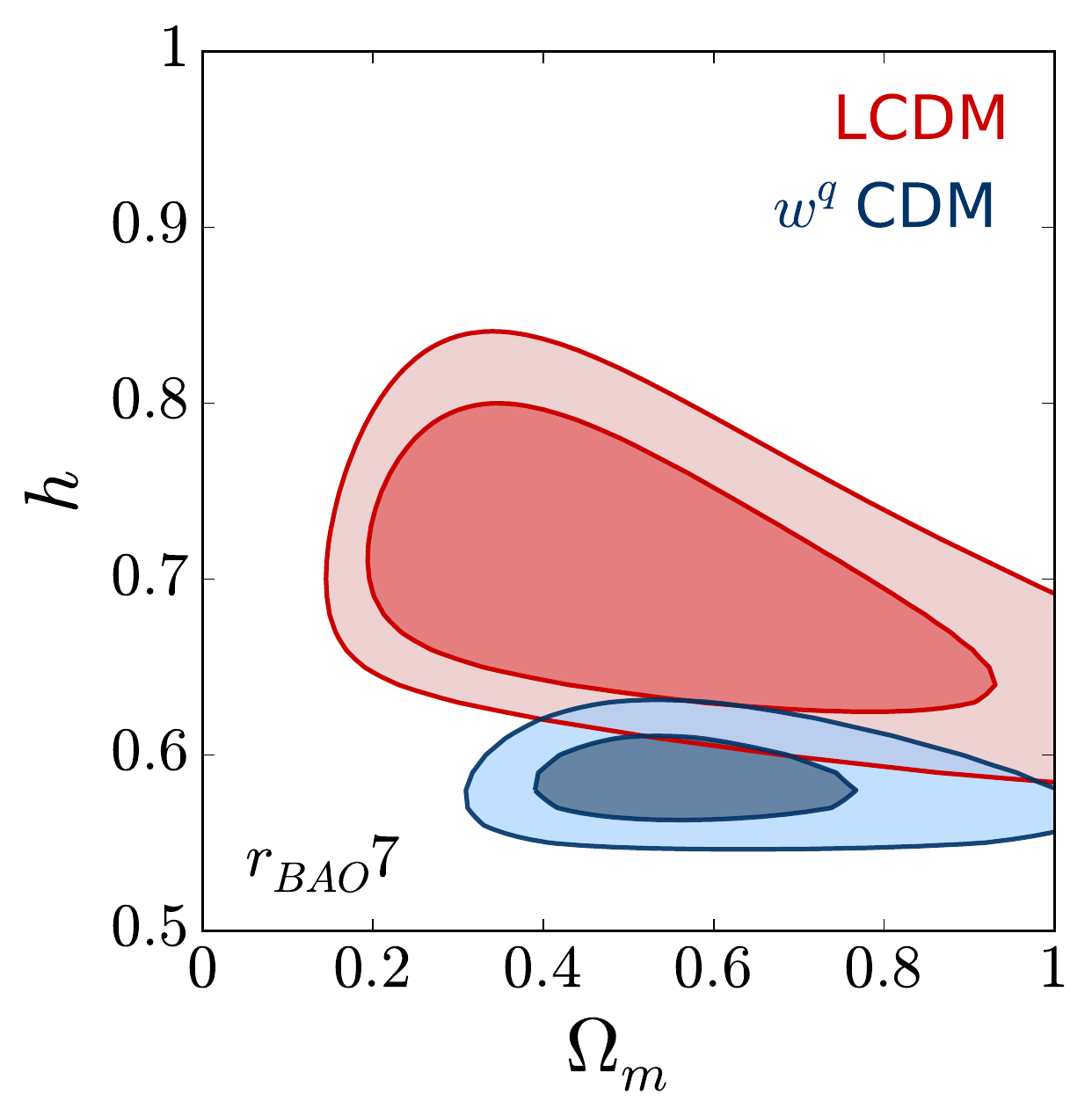}
        \label{fig:BAO3-LCDMFree_7}
        \caption{}
    \end{subfigure}\quad
     \begin{subfigure}[t]{0.39\textwidth}
	\includegraphics[width=0.9\linewidth]{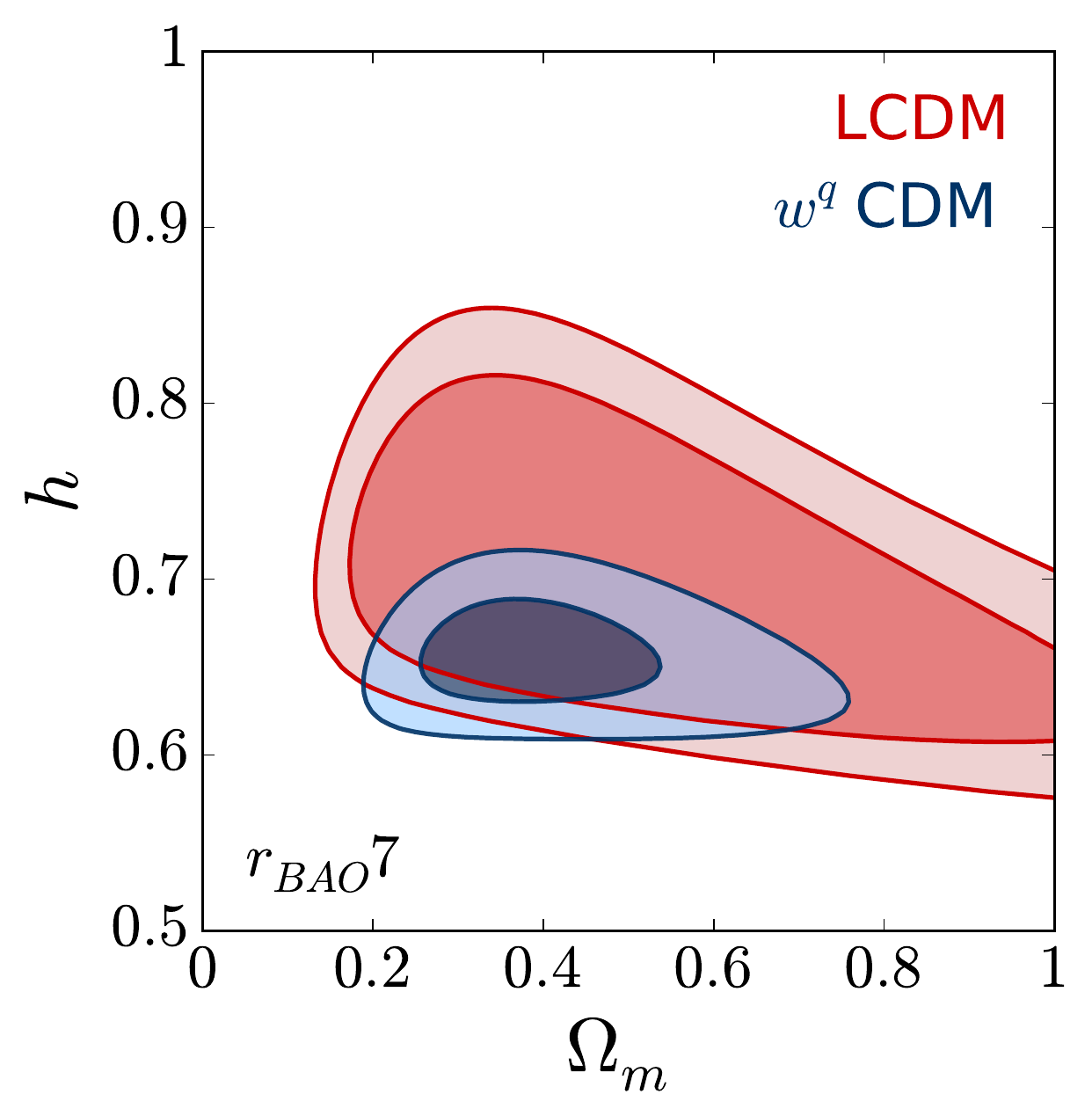}
\label{fig:BAOPlanck2-LCDMPlanck_7}
        \caption{}
    \end{subfigure}\quad

    \centering
		 \begin{subfigure}[t]{0.35\textwidth}
	 \captionsetup{width=5cm}
        \includegraphics[width=\textwidth]{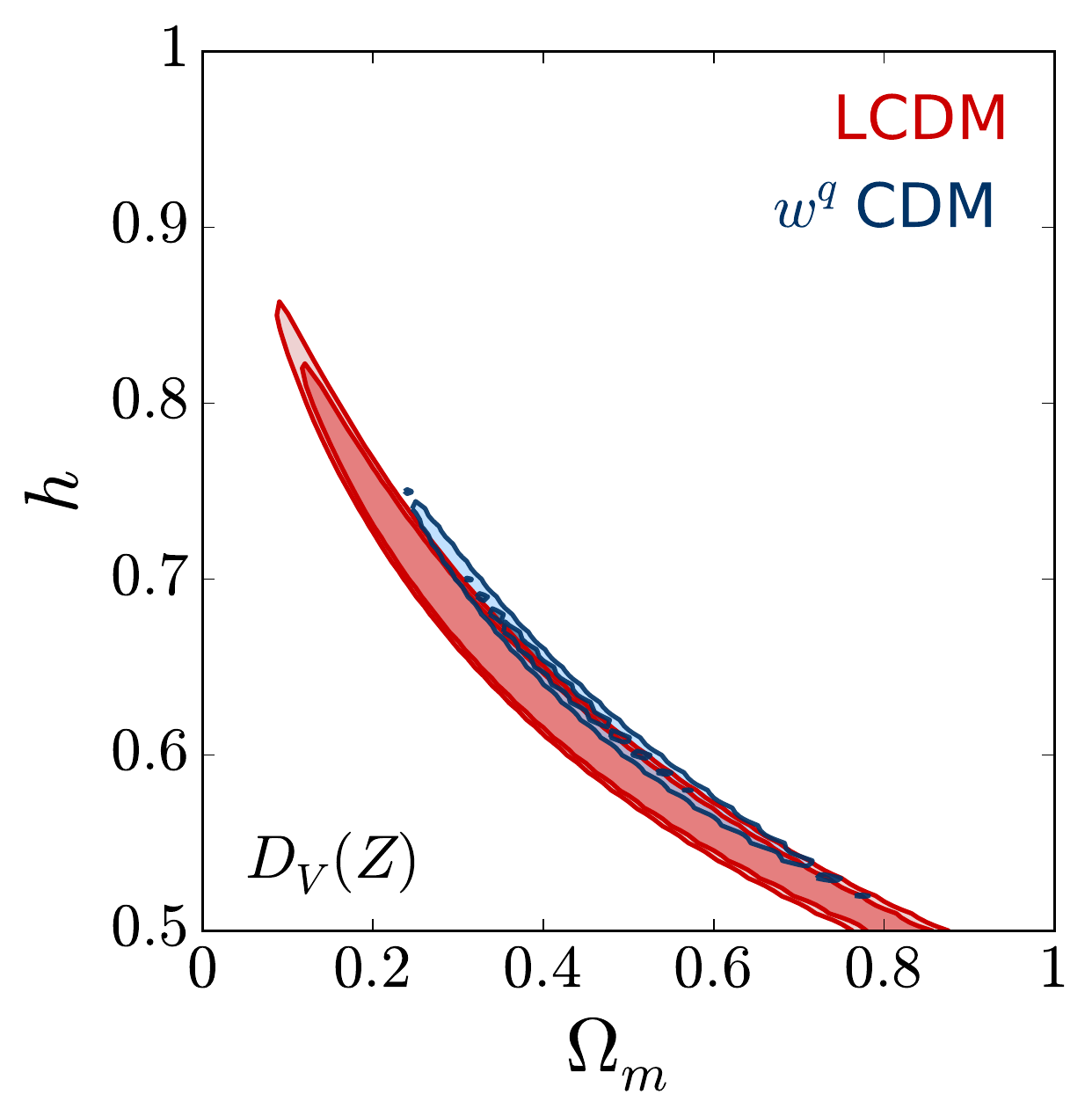}
        \label{fig:BAO3-LCDMFree_7}
        \caption{}
    \end{subfigure}\quad
     \begin{subfigure}[t]{0.39\textwidth}
	\includegraphics[width=0.9\linewidth]{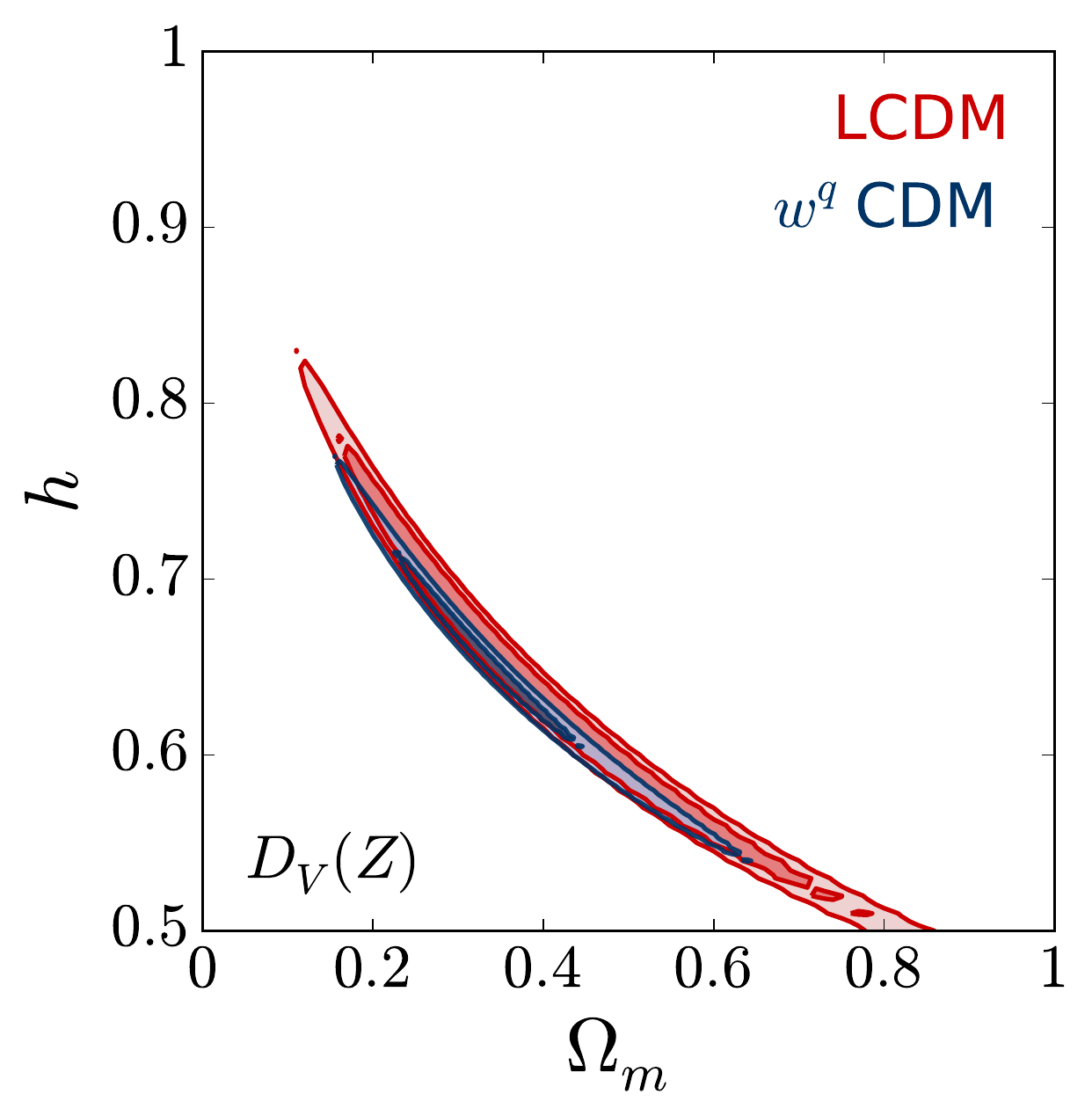}
\label{fig:BAOPlanck2-LCDMPlanck_7}
        \caption{}
    \end{subfigure}\quad

\caption{Figures (a) and (b) show the contours corresponding to $\rbao(z)$ including Ly$\a$-F measurements, (c) and (d) corresponds to $\rbao(z)$ without Ly$\a$-F and (e) and (f) to $D_V(z)$ data. In blue we portrait the contours corresponding to the best fit models and the LCDM contours are shown in red. The left column shows the ``BAO$_{free}$" results and the right one shows the ``BAO + Planck 1$\sigma$" outcomes. }
	\label{fig:contoursLCDM}
\end{figure*}

\begin{figure*}
	\centering
	
	 \begin{subfigure}[t]{0.3\textwidth}
	 \captionsetup{width=4cm}
        \includegraphics[width=\textwidth]{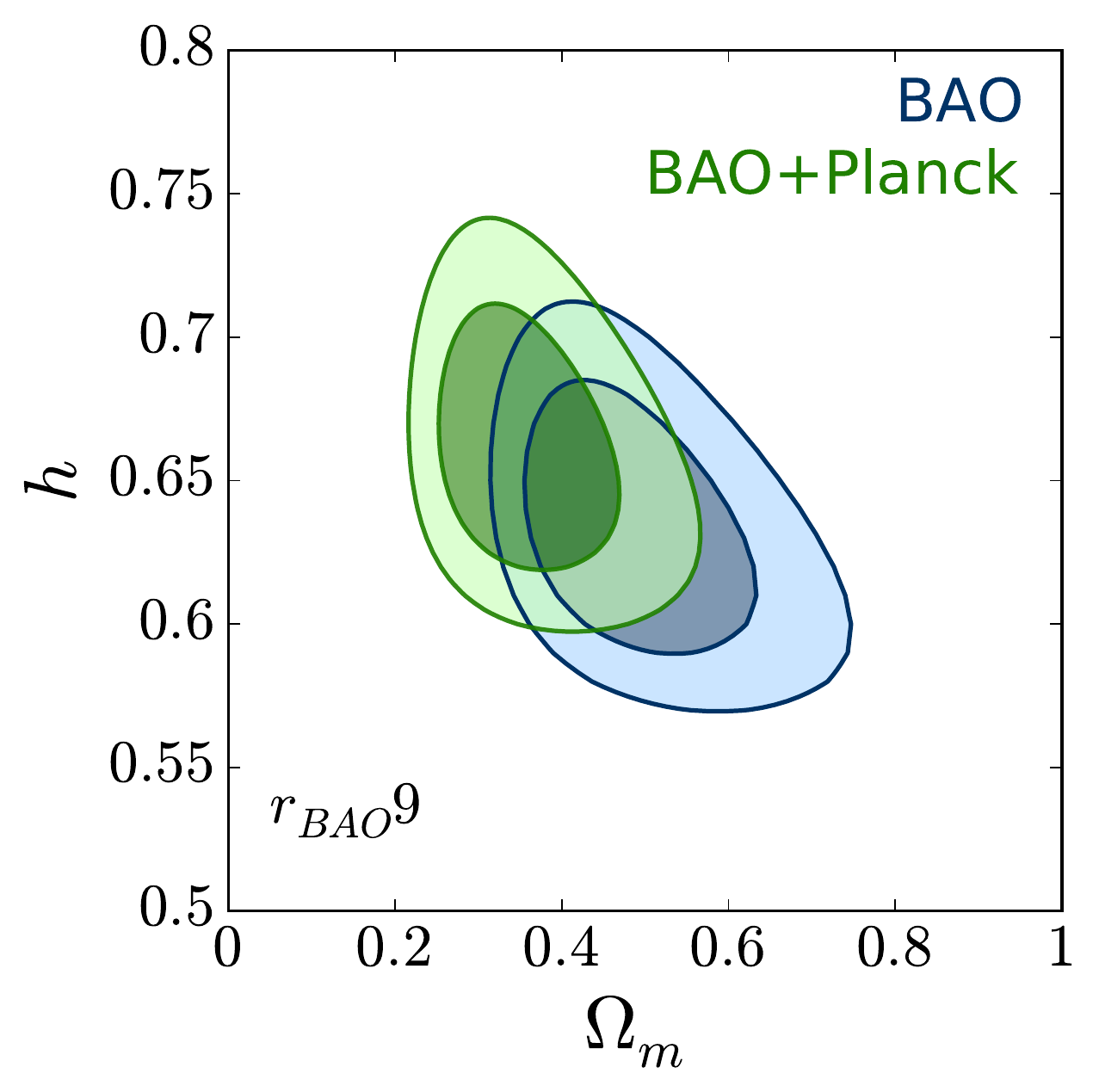}
        \label{fig:BAO-BAOPlanck9}
        \caption{Using the 9 data points of $\rbao(z)$ measurements.}
    \end{subfigure}\quad
    \begin{subfigure}[t]{0.32\textwidth}
	\includegraphics[width=0.9\linewidth]{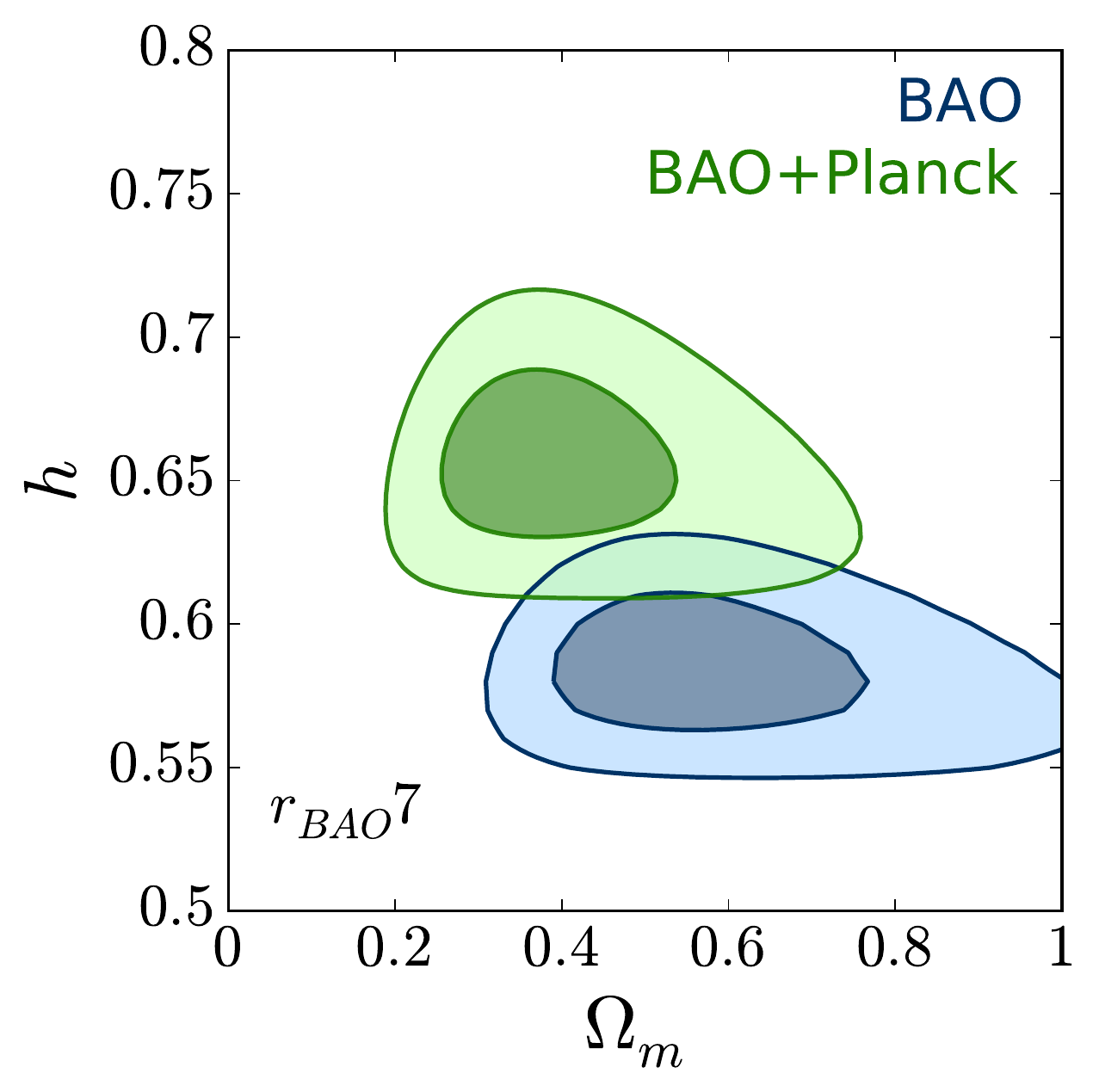}
\label{fig:fig:BAO-BAOPlanck7}
        \caption{Using the 7 data points of $\rbao(z)$ measurements.}
    \end{subfigure}\quad
        \begin{subfigure}[t]{0.32\textwidth}
	\includegraphics[width=0.9\linewidth]{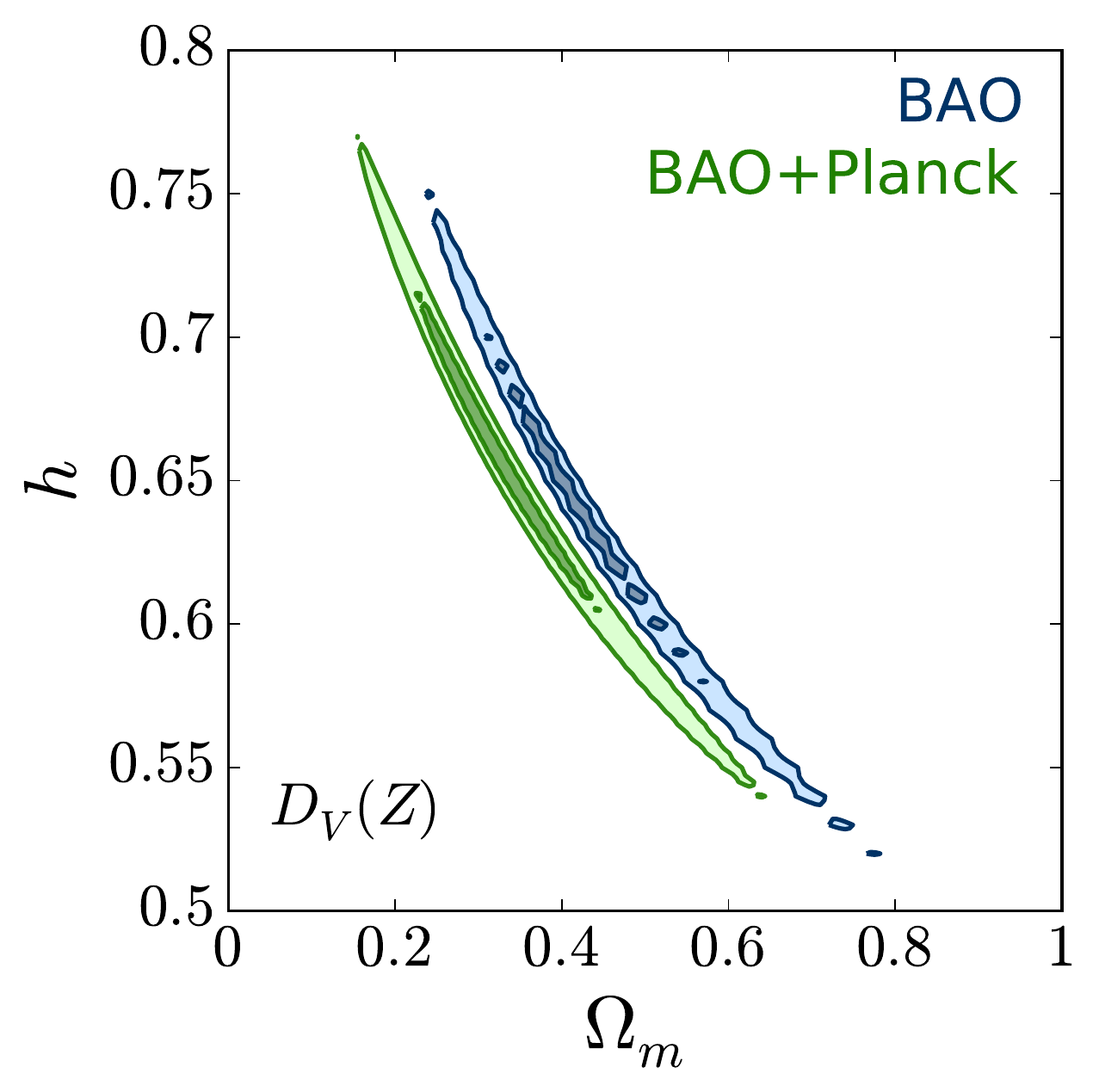}
\label{fig:fig:BAO-BAOPlanckDv}
        \caption{Using the 7 data points of $D_V(z)$ measurements.}
    \end{subfigure}\quad

	\caption{``BAO" contours in blue and ``BAO+Planck" in green. }
	\label{fig:contoursPlanck}
\end{figure*}

\begin{figure*}
	\centering
	
    \begin{subfigure}[t]{0.42\textwidth}
	\includegraphics[width=0.9\linewidth]{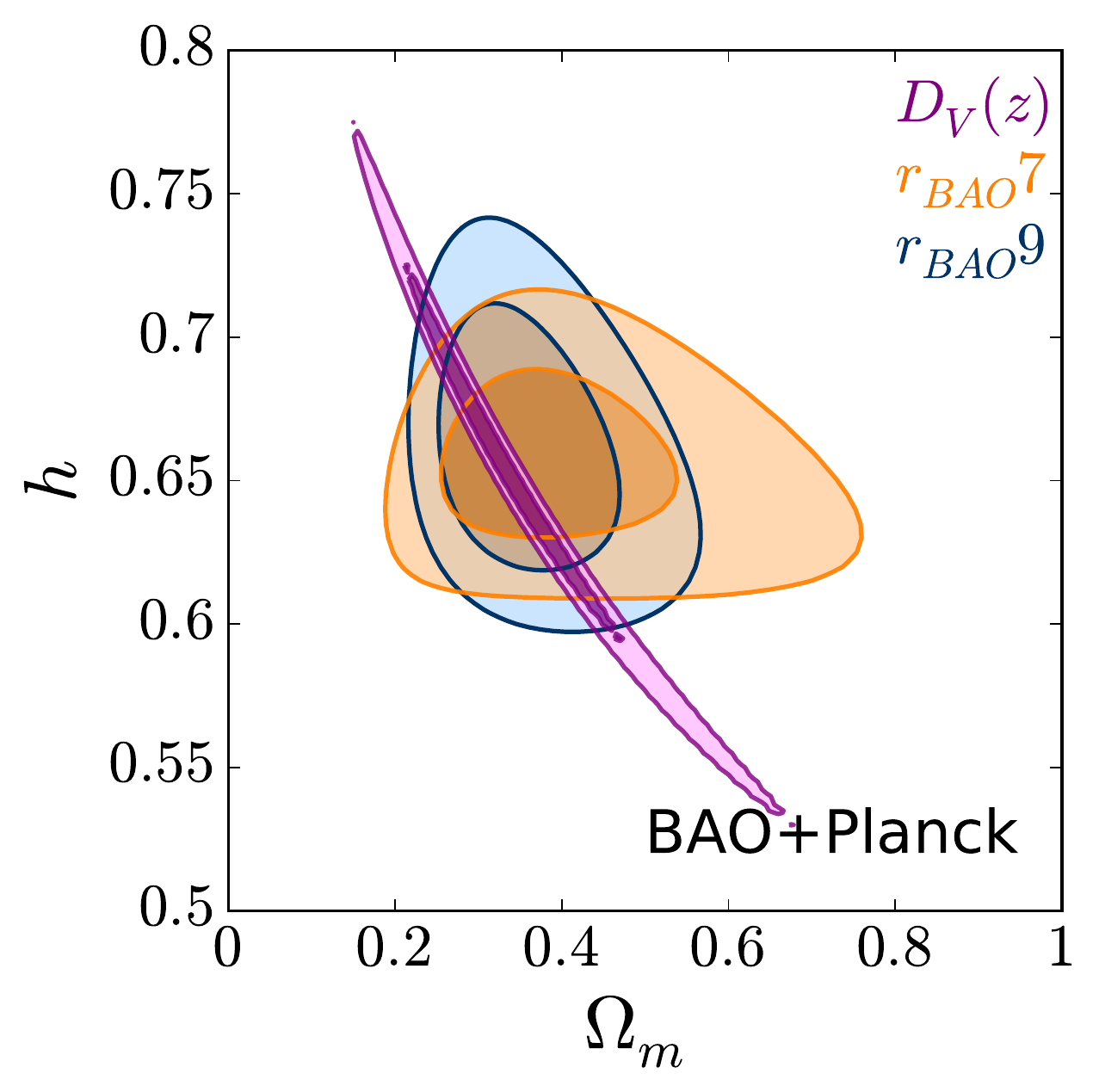}
\label{fig:with_without_LyA_Dv}
        \caption{``BAO+Plank" contours }
    \end{subfigure}\quad
        \begin{subfigure}[t]{0.42\textwidth}
	\includegraphics[width=0.9\linewidth]{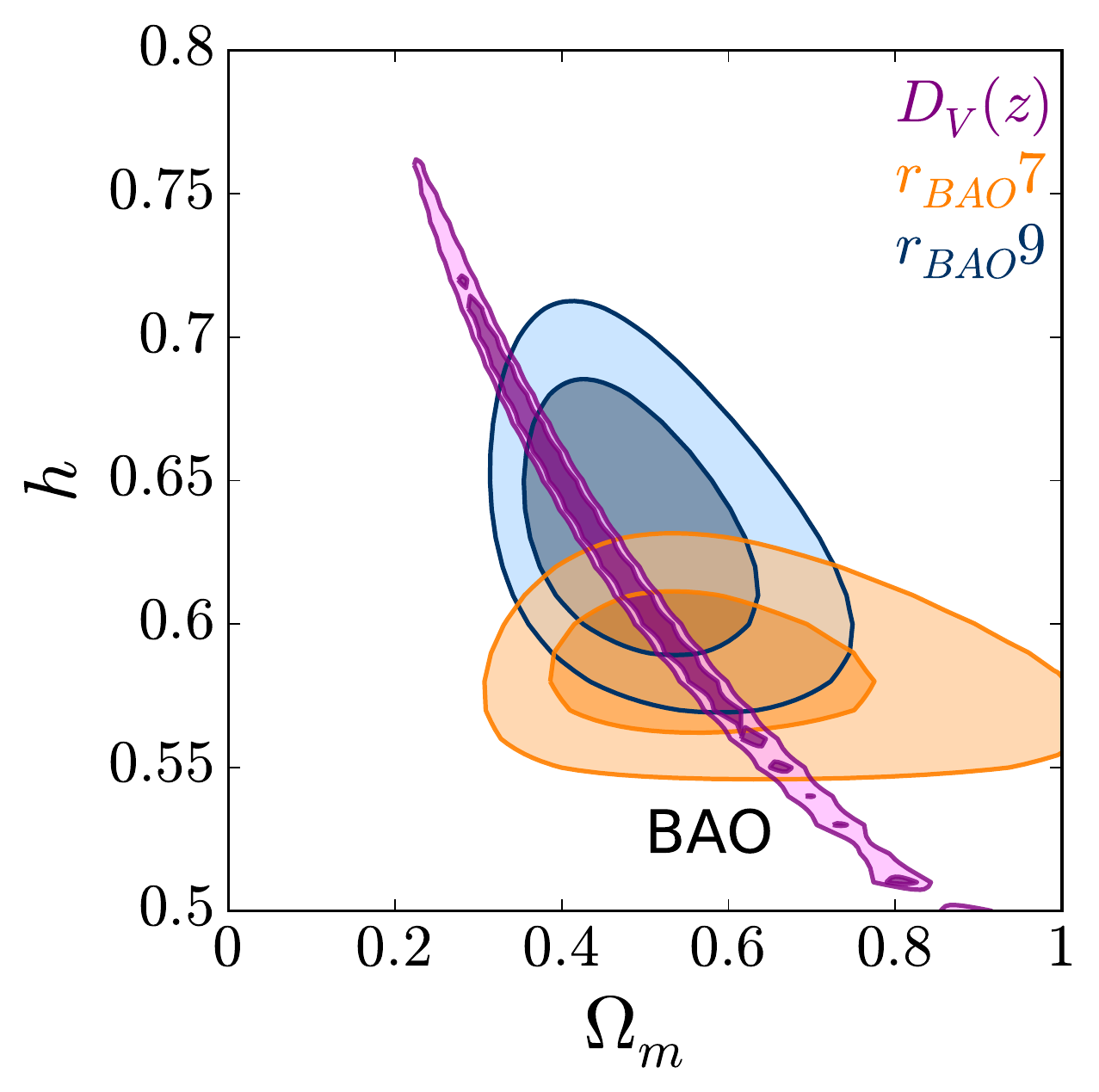}
\label{fig:with_without_LyA_Dv}
        \caption{ ``BAO" contours }
    \end{subfigure}\quad

	\caption{Comparison of the different data sets: $\rbao(z)$ with or without Ly$\a$-F measurements and $D_V(z)$ measurements.}
	\label{fig:contoursDataSets}
\end{figure*}

\begin{figure*}
	\centering
		 \begin{subfigure}[t]{0.45\textwidth}	
		 \captionsetup{width=5cm}
	        \includegraphics[width=\textwidth]{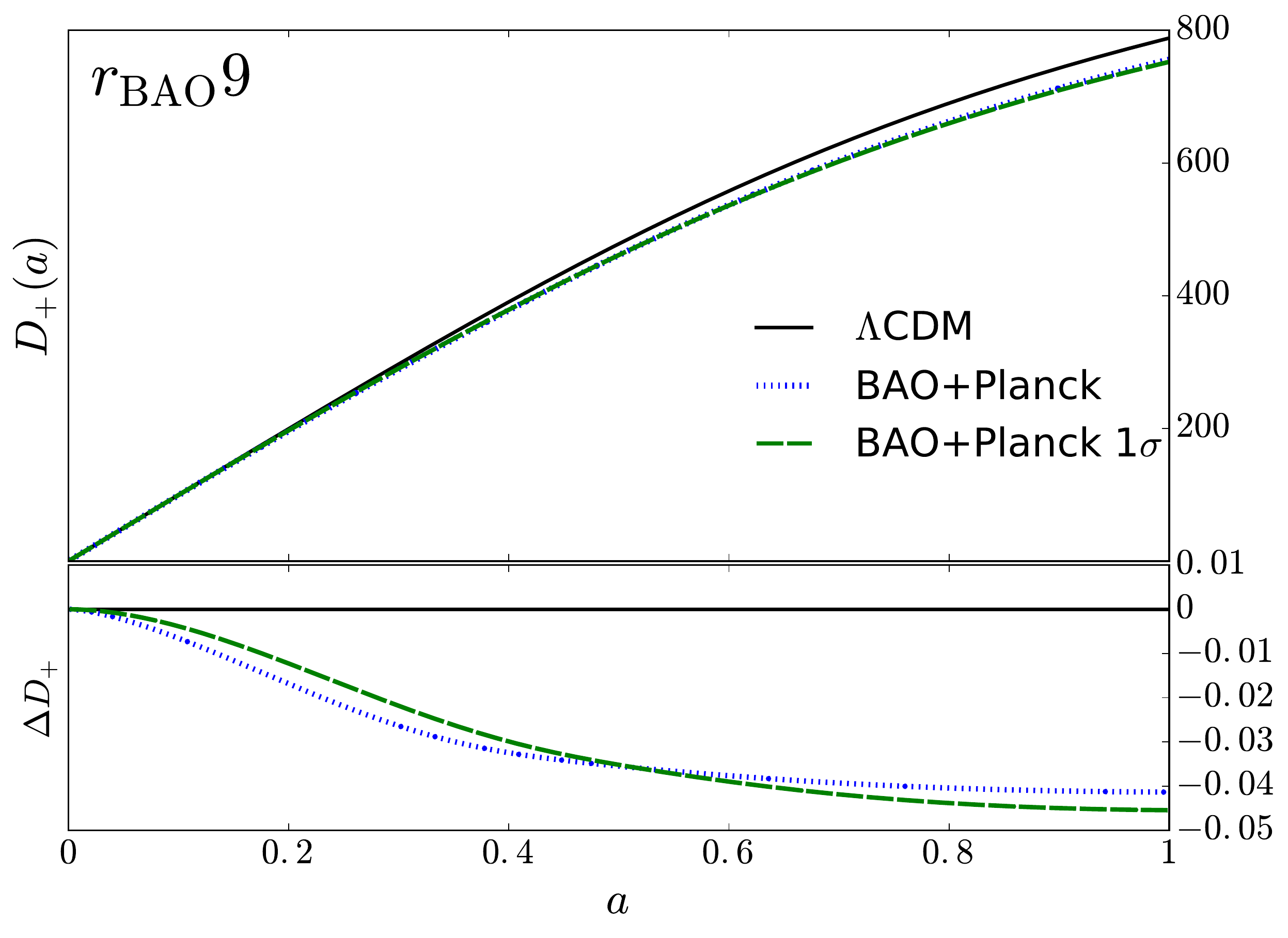}
    \end{subfigure}\quad
     \begin{subfigure}[t]{0.42\textwidth}
	 \captionsetup{width=5cm}
        \includegraphics[width=\textwidth]{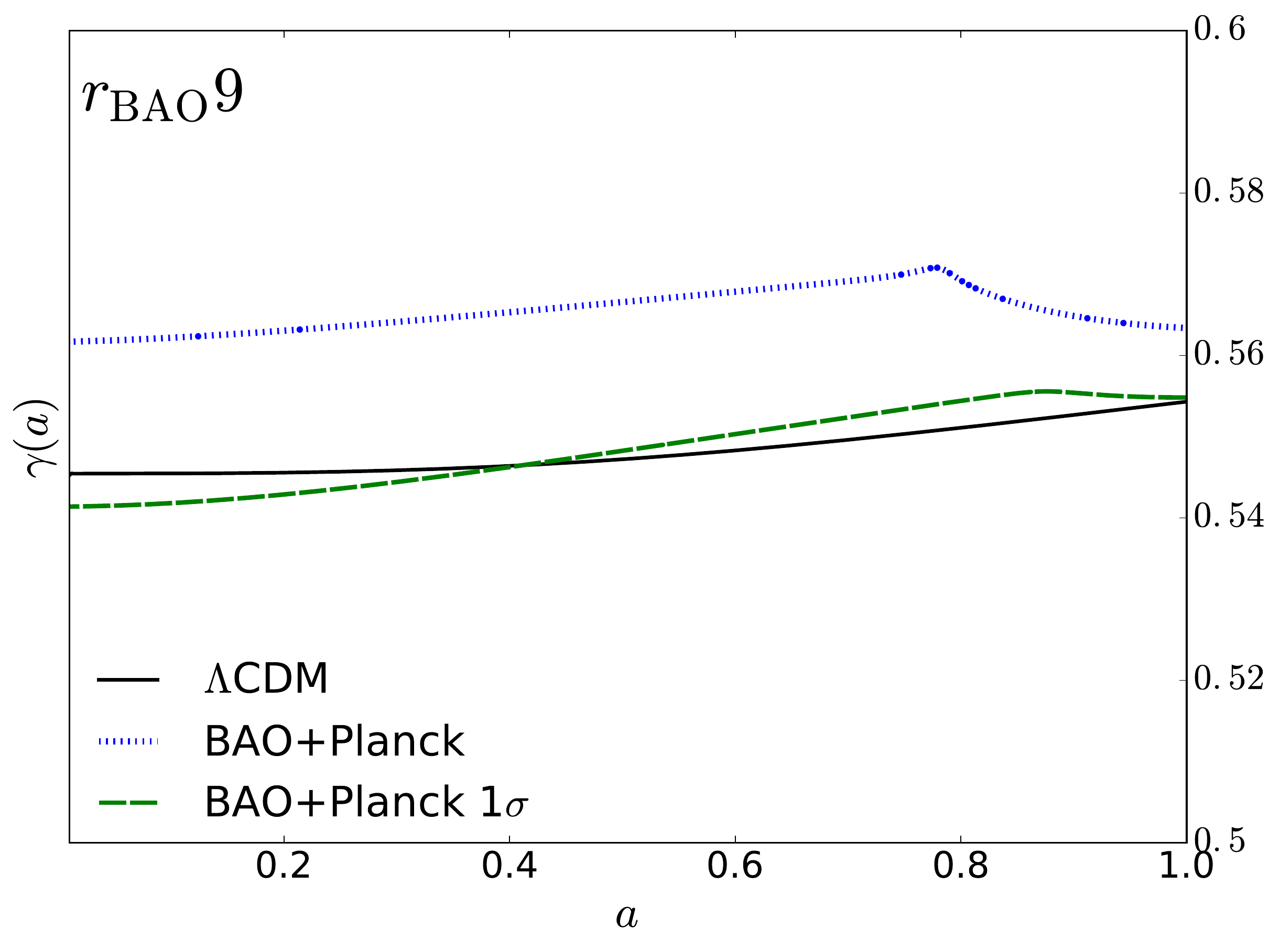}
    \end{subfigure}\quad

     \begin{subfigure}[t]{0.45\textwidth}
	 \captionsetup{width=5cm}
        \includegraphics[width=\textwidth]{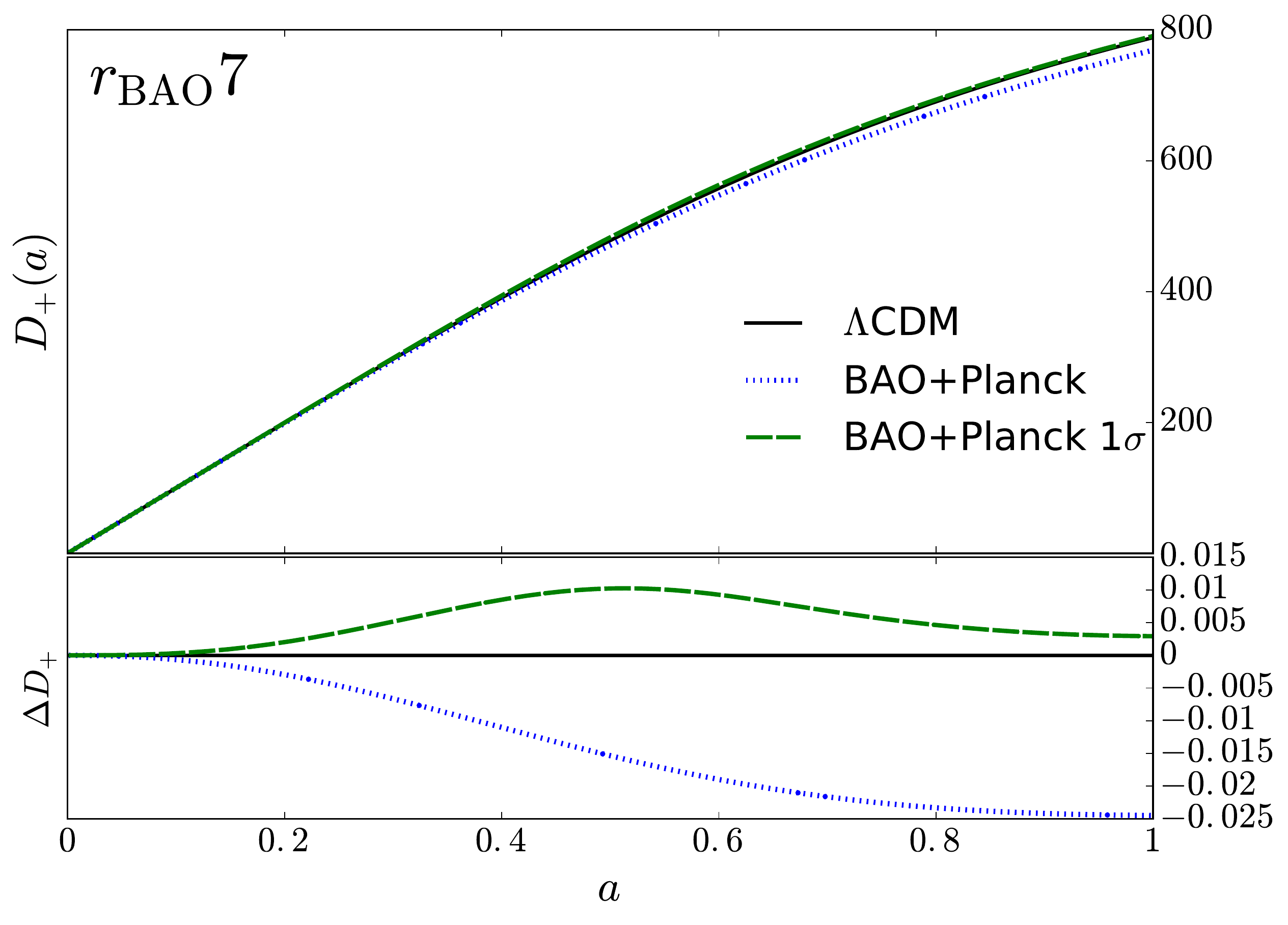}
    \end{subfigure}\quad
   \centering
		 \begin{subfigure}[t]{0.42\textwidth}	
		 \captionsetup{width=5cm}
	        \includegraphics[width=\textwidth]{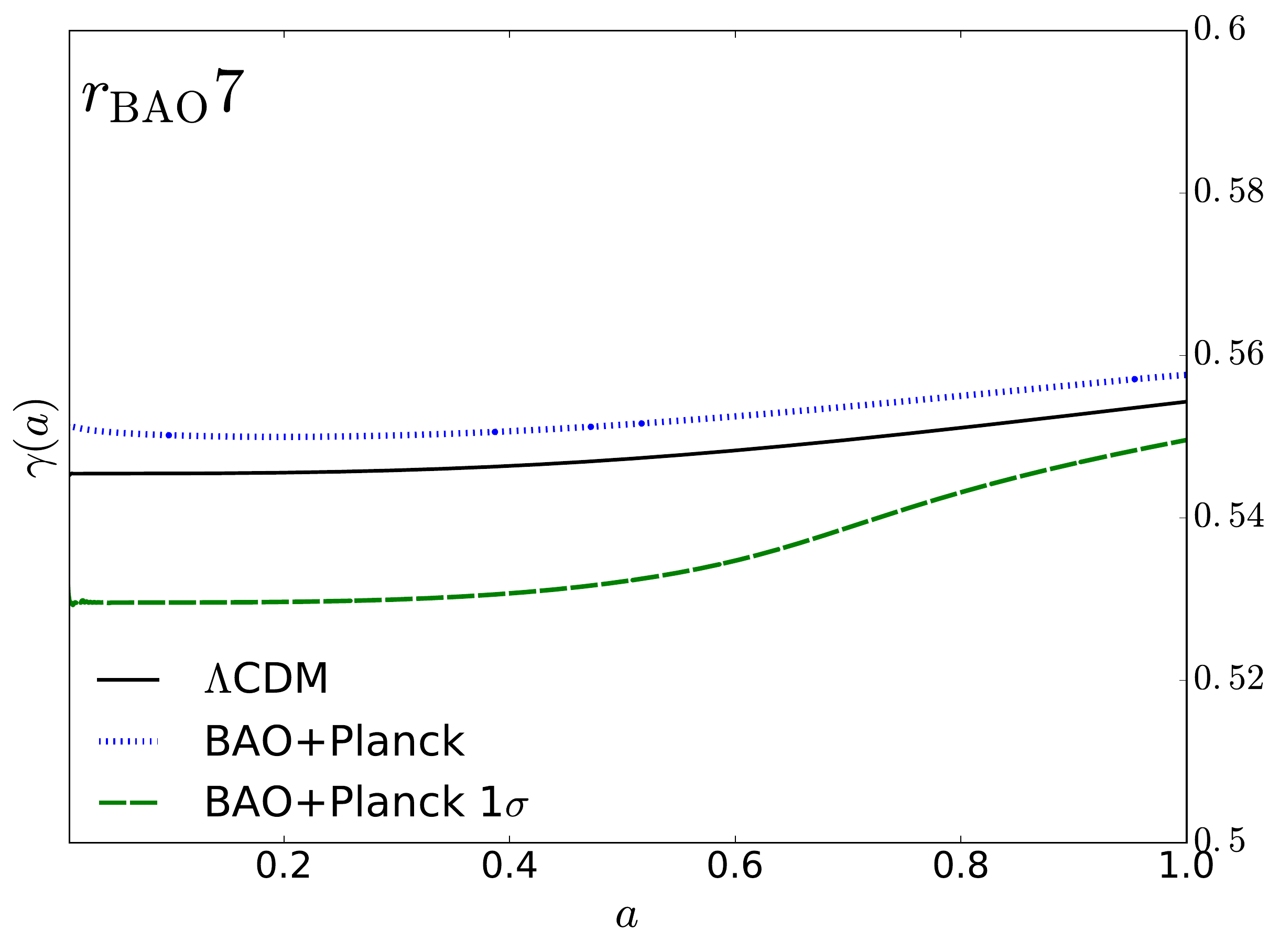}
    \end{subfigure}\quad

     \begin{subfigure}[t]{0.45\textwidth}
	 \captionsetup{width=5cm}
        \includegraphics[width=\textwidth]{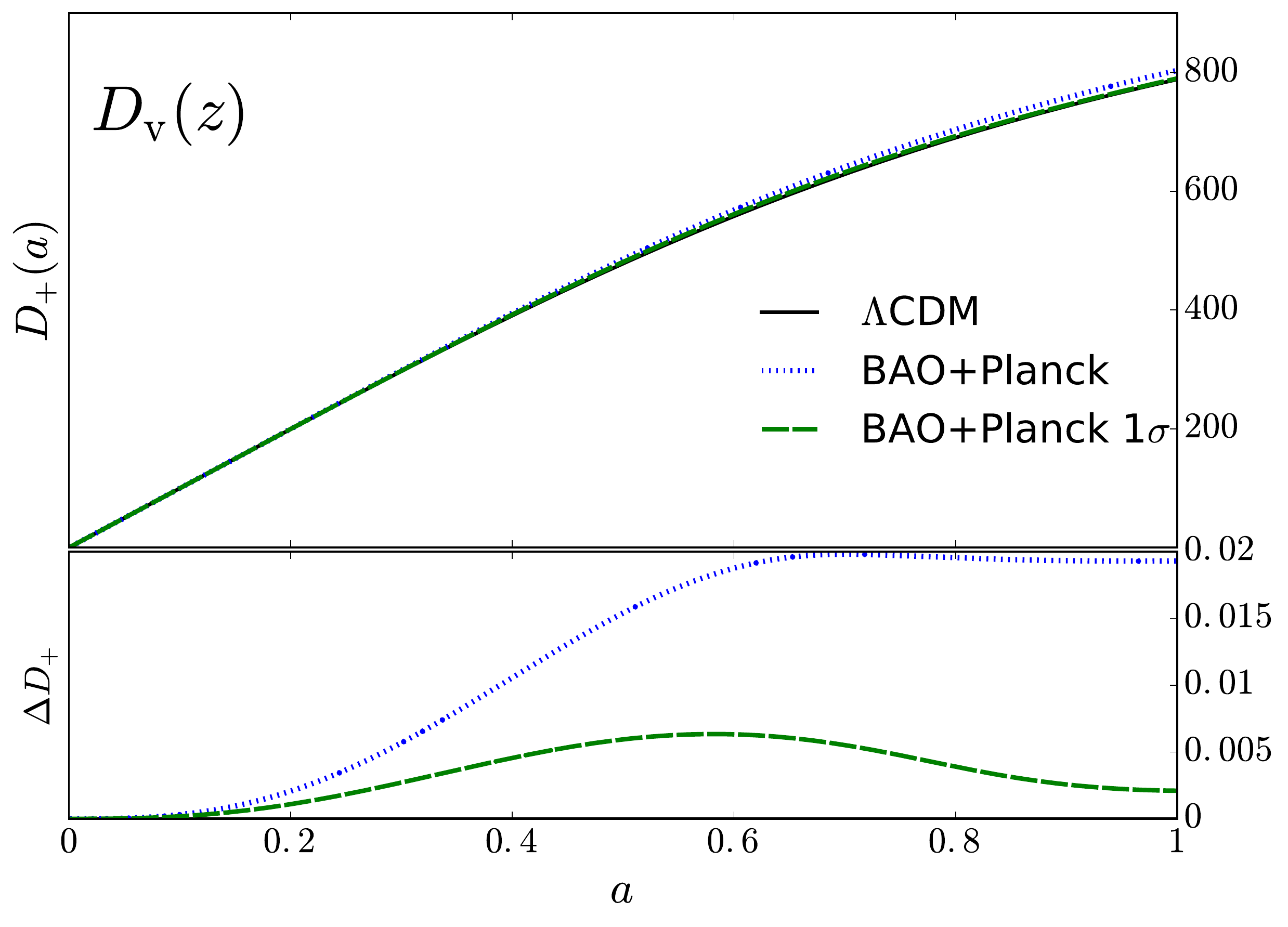}
             
        \caption{Growth function, $D_{+}(a)\equiv \delta_m(a)/\delta_m(a_{ini})$ for the different DE  models. The bottom panel displays the comparison to the growth when a cosmological constant is assumed. }
    \end{subfigure}\quad
     \begin{subfigure}[t]{0.42\textwidth}
	 \captionsetup{width=5cm}
        \includegraphics[width=\textwidth]{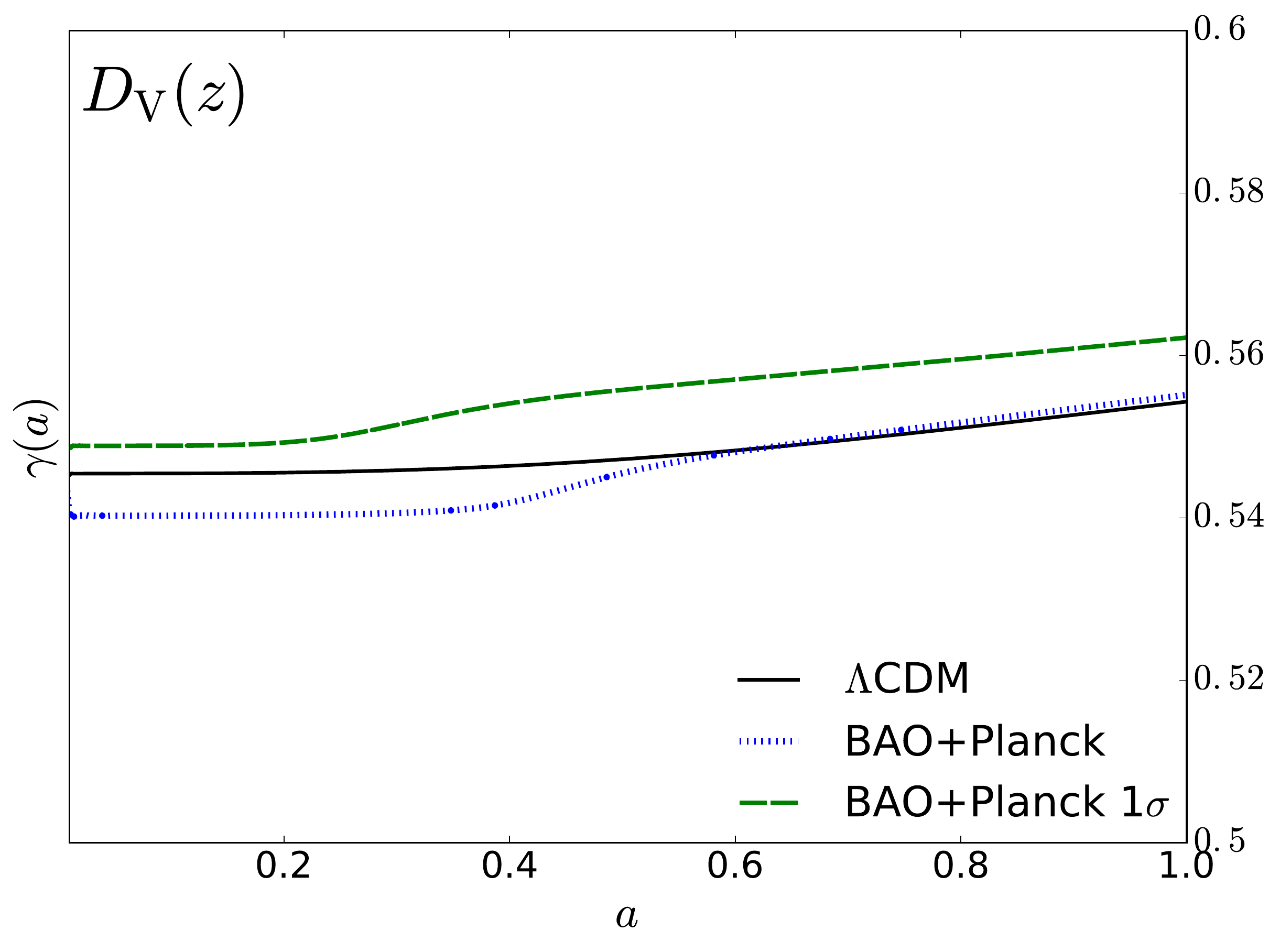}
        \caption{Growth index, $\gamma(a)$,  for the different DE  models, with the ansatz $f(a)\equiv\frac{dln\delta_m(a)}{dlna}= \Omega_m(a)^{\gamma}$.}
    \end{subfigure}\quad
    \caption{Growth of matter perturbations when assuming background Dark Energy with a state equation given by a cosmological constant or the results labeled  ``BAO + Planck" and ``BAO + Planck 1$\sigma$" as reported in Tables \ref{table:results9pts}, \ref{table:results7pts} and \ref{table:resultsdvz}, respectively.}
         \label{fig:perturbations}
\end{figure*}

Additionally to the numerical minimization, a direct comparison of our results and the fiducial model ($\La$CDM  with the central values for $\Om$ and $h$ from Planck) was made. In order to do so we computed the ratio of the functions $\rbao$ and $D_V(z)$ when assuming each one of the best fit models resulting from the minimization to the corresponding value when the fiducial model ($\Lambda$CDM) is assumed. We report the results from that analysis in Figure \ref{fig:ratios}. The curves were plotted on top of the observational data with the associated relative error of the measurements. The gray bands were estimated by analyzing the change on the fiducial function when the cos\-mo\-lo\-gi\-cal parameters were changed by 1$\s$, twice and three times that quantity from their central values.



\section{Results}
\la{sec:Results} 

Table \ref{table:results9pts} shows the results when we use  $\rbao(z)$  with Ly$\a$-F data (hereafter  referred to as ``$\rbao$ 9" indicating the nine data points in that set) . When we exclude the Ly$\a$-F data points (``$\rbao$ 7"), the corresponding best fit values are those listed in Table \ref{table:results7pts} and finally, when we we used the measurements for $D_V(z)$, the results are those reported in Table \ref{table:resultsdvz}. The corresponding EoS curves are in Figure \ref{fig:bfm} and the ratio plots are shown in Figure \ref{fig:ratios}.

From Table \ref{table:results9pts} we see that the $\chi^2$ value obtained for the parametrization \ref{eq:eos} with the five different  priors\- used is much smaller that the one in the $\L$CDM case and the $\chi^2_{red}$ value is smaller than unity for all the  results presented.  The value for $w_0$ when we combined BAO and Planck data, lies very close to $-1$ and the values for the transition redshift obtained were large  ($z_T>1.16$). The models follow a dynamic of the  type called ``freezing" (\cite{Caldwell:2005tm}, \cite{Linder:2007wa}, \cite{Huterer:2006mv}, \cite{Linder:2006sv}) given the central values for  $w_i>-1$ (although the $\pm 1 \sigma$ error limits  allow for $w_i\in$[-2, 0] for the model ``BAO + $Planck_{\Omega_m}$"). 
When only the BAO data was used, $w_0=-1$ is excluded with 1$\s$ of statistical significance, we got a central value for $w_i> -1$ implying a dynamics of the type calle ``thawing" (\cite{Caldwell:2005tm}, \cite{Linder:2007wa}, \cite{Huterer:2006mv}, \cite{Linder:2006sv}),  the transition redshift value is  smaller ($z_T\geq0.37$) and the fractional matter density preferred is larger than $\overline{\Omega}_m$.  The scenario $q=z_T=1$ is allowed by the  $1\sigma$ limits for two models (``BAO + Planck" and ``BAO + Planck 1$\s$"). 
The value for $\chi^2$ increases with tighter constraints used in every prior, as to be expected. 

From Table \ref{table:results7pts} we see that parametrization \ref{eq:eos} gets a much smaller value of the $\chi^2$ function than that of $\Lambda$CDM, but the $\chi_{red}^2$ value is not smaller than unity for all  the models.  The value $w_0=-1$ is excluded at $1\s$ of statistical significance for all but one model (``BAO + Planck ").  We do not find a clear separation of behaviors (freezing vs thawing) in the models, and we interpret this as a result of the much limited range of available data ($0.106\geq z\leq 0.73$); the central value for the parameter $w_i$ is $w_i<-1$ in all the models. The scenario $q=z_T=1$ is excluded with 1$\s$ of statistical significance in all but two models: ``BAO$_{\Omega_m}$" and ``BAO+Planck$_{\Omega_m}$". 

One thing in common for these two tables  is that, the larger the value for $\Omega_m$, the smaller the value for $w_0$ and the smaller the value for $z_T$.

The models summarized in Table \ref{table:resultsdvz} have a $\chi^2$ smaller than that for $\Lambda$CDM although the difference is not as big as in the other two tables. We find an inverse behavior in the relationship $\Omega_m-w_0$ than in the previous results: for this models, the smaller the value for $\Omega_m$, the smaller the value for $w_0$. Again, no clear  distinction between ``freezing" and ``thawing" dynamics was found and the $w_0=-1$ case is included at $1\s$ only for the models ``BAO+Planck" and ``BAO+Planck$_{\Omega_m}$".  The case when $q = z_T=1 $ is excluded by all the models except for the ``BAO + Planck" one.

In the set of results summarized in Tables \ref{table:results9pts}-\ref{table:resultsdvz}, we find that while the value for $w_0$ was fairly restricted, the data seems not to be very sensitive  to  the rest of the free parameters, in particular to the values for $w_i$ and $z_T$ within the tested range  of those two parameters (Table \ref{table:boundaries}).  

In Fi\-gu\-re \ref{fig:w0-zT-q} we show the $1\s$-$2\s$ confidence regions for the $w_0$-$z_T$ and $w_0$-$q$ spaces when  the rest of the parameters were fixed to their best fit values assuming the models labeled as ``BAO + Planck 1$\s$". The plots for $q$-$z_T$ represent the contours of $\chi^2$ around the minimum when the rest of parameters were fixed to their best fit values. 

We understand this insensitivity to the value of the steepness of the transition as due to the short number of data points, which are not enough to fix the slope of the transition while the transition redshift  cannot be too precisely fixed due to the  limitation in the range of the measurements ($ z \leq 2.36$). With the increase in the number and range of future measurements, better constraints will be obtained.

\begin{table*}	
	\begin{center}
	\begin{tabular}{|c|c|c|c|c|c|c|c|c|c|}
	\hline
	\multicolumn{10}{|c|}{$r_{BAO}$ 9 }\\
	\hline \hline
	Alias   & $\chi^2$ &  $\chi_{red}^2$ &$\nu$=9-M& $w_0$ & $w_i$& q  &$z_T $&$\Om$ & h \\
	\hline \hline
	$\La CDM_{Fid}$ & 7.29 &  0.81& 9 & -1 & -1 & 1 & 1 & $\overline{\Omega}_m$ & $\overline{h} $ \\
	 \hline

	$BAO_{\Om}$ & 2.44 &  0.61 & 4 &-$0.88^{+0.11}_{-0.12}$ & -1.82($\leq$-0.79) & 7.77($\geq$1.60) & 0.50($\geq$0.35) & 0.4173  & $\overline{h}$ \\
	
	$BAO_{free}$ & 2.35 & 0.78& 3  &$-0.65^{+0.11}_{-0.14}$   & -2.0($\leq$-1.52) & 4.87($\geq$1.55) & $0.37^{+0.09}_{-0.08}$ &  0.4934 & 0.64 \\
	
	\hline
	BAO+Planck & 3.10 &  0.62 & 5 &  $-0.95^{+0.12}_{-0.07}$ & -0.51($\leq$0.0) & 9.21($\geq$0.42) & 1.38 ($\geq$0.34) & $\overline{\Omega}_m$ & $\overline{h}$  \\
	
	$BAO+Planck_{\Om}$  & 2.94 & 0.74  & 4 & $-0.95^{+0.10}_{-0.07} $&0($\geq$-2.0) & 2.94($\geq$0.9) &  1.92$(\geq$0.86) & 0.3171 & $\overline{h}$  \\
	
	BAO+Planck 1$\s$ & 2.85 &  0.95 & 3  & $-0.91^{+0.06}_{-0.08}$ & -0.62($\leq$0.0) & 9.95($\geq$0.56) & 1.16($\geq$0.36) & 0.3247 & 0.67\\
	 \hline
	\end{tabular}
	\caption {Best Fit Values for the fitted parameters when we used data points of $\rbao$ including Ly$\a$-F measurements. $\chi^2$ represents the minimum value for the chi-squared function, $\chi_{red}^2 = \chi^2/\nu$ is the reduced chi-squared function where $\nu$ stands for the free degrees of freedom and $M$ is the number of parameters to be constrained.  The free parameters of the proposed EoS, $w_0$, $w_i$, $q$,  $z_t$, are reported with their respective 1$\s$ error limits. $\overline{\Omega}_m$ and $\overline{h}$ represent the central value for $\Omega_m$ and $h$, respectively.}
	\la{table:results9pts}
	\end{center}
\end{table*}

\begin{table*}	
	\begin{center}
	\begin{tabular}{|c|c|c|c|c|c|c|c|c|c|}
	\hline 
	\multicolumn{10}{|c|}{$r_{BAO}$ 7 }\\
	\hline \hline
	Alias & $\chi^2$ & $\chi_{red}^2$& $\nu$=7-M& $w_0$ & $w_i$& q  &$z_T$  &$\Om$  & $h$  \\
	\hline \hline
	$\La CDM_{Fid}$ & 7.29 &1.04 & 7 & -1 & -1 & 1 & 1 & $\overline{\Omega}_m$ & $\overline{h}$   \\
	
 	\hline
	$BAO_{\Om}$ & 2.53 &1.26 & 2 & $-0.88^{+0.06}_{-0.10}$ & $-1.16^{+0.40}_{-0.81}$ & 4.62($\geq$0.24)& 0.44($\geq$ 0.19) & 0.3664 & $\overline{h}$  \\
	
	$BAO_{free}$ & 2.21 &2.21 & 1 &  $-0.24^{+0.12}_{-0.07}$   & -1.80($\leq$-1.59) &  2.48($\geq$1.57)& $0.24^{+0.05}_{-0.02}$&  0.6408 & 0.58  \\
	
	\hline
	BAO + Planck & 2.73 &0.91 & 3 &  $-0.91^{+0.08}_{-0.09}$ & -2.0($\leq$0.0) & 7.16($\geq$2.3) & 0.69($\geq$0.47)& $\overline{\Omega}_m$ & $\overline{h}$ \\
	
	$BAO+Planck_{\Om}$  & 2.73 &1.36 & 2 & $-0.89^{+0.06}_{-0.08}$ & -1.36($\leq$-0.49) & 2.77($\geq$1.0) &  0.67($\geq$ 0.37) & 0.3247 & $\overline{h}$\\

	BAO + Planck 1$\s$ & 2.71 &2.71& 1 & $-0.89^{+0.03}_{-0.06}$ & -1.28($\leq$-0.56)& 4.65($\geq$1.34) & 0.64($\geq$ 0.38)& 0.3245 & 0.67  \\
	 \hline
	\end{tabular}
	\caption{Best Fit Values for the fitted parameters when the $r_{BAO}(z)$ data points were used but without Ly$\a$-F measurements.  }
	\la{table:results7pts}
	\end{center}
\end{table*}	

\begin{table*}
	\begin{center}
	\begin{tabular}{|c|c|c|c|c|c|c|c|c|c|}
	\hline
	\multicolumn{10}{|c|}{$D_{V}$(z)}\\
	\hline \hline

	Alias & $\chi^2$ & $\chi_{red}^2$&$\nu$=7-M & $w_0$ & $w_i$& q &$z_T$  &$\Om$ &$h$  \\
	\hline \hline
	$\La CDM_{Fid}$ & 2.62 &0.37 & 7 & -1 & -1 & 1 & 1 & $\overline{\Omega}_m$ & $\overline{h}$  \\

	\hline
	
	$BAO_{\Om}$  & 1.95 & 0.97 & 2 &  $-0.49^{+0.03}_{-0.11}$ & -1.56($\leq$-1.2) & $3.74^{+2.36}_{-1.74}$ & 0.56$^{+0.10}_{-0.11}$ & 0.2084 & $\overline{h}$ \\
	
	$BAO_{free}$ &1.62 & 1.62 & 1  &   $-0.44^{+0.03}_{-0.04}$   & $-0.94^{+0.13}_{-0.14}$ & 5.25($\geq$2.1) & 0.37$^{+0.07}_{-0.04}$ &  0.1624 & 0.69  \\ \hline
	
	BAO+Planck & 2.57 & 0.85& 3 &$-0.96^{+0.09}_{-0.10}$ & $-1.23^{+0.60}_{-0.64}$ & 9.35($\geq$0.3) & 0.41($\geq$0.20) & $\overline{\Omega}_m$ & $\overline{h}$  \\
	
	$BAO+Planck_{\Om}$  & 2.55 & 1.27& 2 & $ -0.94\pm0.07 $& -1.99($\leq$-0.63) & 9.91 ($\geq$3.2)&  0.51($\geq$0.44) & 0.3133 & $\overline{h}$  \\
	
	BAO+Planck  1$\s$ & 2.43  & 2.43& 1 & $-0.87^{+0.07}_{-0.05}$ & $-1.10^{+0.17}_{-0.12}$ & 4.94 ($\geq$0.3)&  $0.27^{+0.26}_{-0.08}$& 0.3122 & 0.67  \\
	 \hline
	\end{tabular}
	\caption{Best Fit Values for the fitted parameters when the  $D_V(z)$  data were used.}
	\la{table:resultsdvz}
	\end{center}
\end{table*}

From Figure \ref{fig:data&curves} we can see that all of our fits lie within the observational errors of the data points.

In  figure   \ref{fig:ratios}  we compared the obtained models to the corresponding $\La$CDM model when only the information of the full CMB was taken into account (taking the fiducial model to be $\La$CDM with Planck + TT+ TE+ EE+low P \ci{Ade:2015xua}).  The observational data points were also divided by the corresponding $\La$CDM pre\-dic\-tion and the error bars were constructed taking the relative error of each measurement. When the data sets of $\rbao(z)$ were used (Figures (a) and (b) of \ref{fig:ratios}), neither the curves of the best fit models nor the $\La$CDM pre\-dic\-tion agreed with the $z = 0.32$ measurement (\ci{Anderson:2013oza}). 
The curves shown in the second panel of Figure \ref{fig:ratios} are contained within the observational error of the data points (except of the $z=0.32$ measurement, as already discussed). The ``$BAO_{free}$" curve agrees only with $z=0.106$, $z=0.44$ and $z=0.73$ data points.

When the $D_V(z)$ data points were used instead, all the curves were contained within the observational data error bars, except for the point corresponding to the $z=0.15$ (\ci{Ross:2014qpa}), which is only in agreement with the ``BAO$_M$" curve.  However, the results derived from this figure depend on the fiducial model chosen (Planck TT+TE+EE + lowP data from \cite{Ade:2015xua}).

To make a further analysis of  the agreement between our parametrization and the $\La$CDM model, we plotted the confidence curves around their minimum for both models. Those results are included in Figure \ref{fig:contoursLCDM}. Generally speaking, when we restricted the values of $\Om$ and $h$ to be within the boundaries of Planck, we see that the two set of contours are very compatible, which is not quite the case when  we remove the boundaries and let $\Om$ and $h$ to vary freely. 

In Figure \ref{fig:contoursPlanck} we analyzed explicitly the difference between adding or not the information of the CMB through the boundaries ($\pm 1 \s$) from Planck. We note that without  the inclusion of Planck boundaries, the data  prefers a higher  $\Om$ value and lower $h$.  In figure \ref{fig:contoursDataSets} we show the contours obtained by using different data sets both when including the Planck boundaries and  letting $\Om$ and $h$ to vary freely. We can see that the data sets are complementary in the sense that, adding the Ly$\a$-F points reduces the contours around the mi\-ni\-mum, meaning that the constraints are tighter and the $D_V(z)$ contours are also included by the $\rbao(z)$ curves.

The analysis of the perturbative regime for different DE models is a promising way for a better understanding of the cause of acceleration in the cosmic expansion. We will study  the DE perturbations  for the parametrization here  introduced and its effect on the growth of structure in a next paper \cite{Macorra:jaber2016perturbs}.

\subsection*{Constant equation of state or CPL parametrization}

The resulting constraints for an  equation of state  assumed to be either a constant, $w$, or given by the CPL parametrization, $w(a)=w_0+(w_0-w_i)(1-a)$, (\ci{Chevallier:2000qy}, \ci{Linder:2002et}) are  displayed in Tables \ref{table:w-cpl-9pts}-\ref{table:w-cpl-dvz}.  The value of both  $h$ and $\Om$ were kept fixed to the central values of Planck (table 4 of \ci{Ade:2015xua}).

For simplicity, in Tables \ref{table:w-cpl-9pts}-\ref{table:w-cpl-dvz}  we denoted by $w_0$ both the value of the equation of state when it was assumed to be constant and also the present value of $w(z)$ when the CPL parametrization (equation  \ref{eq:CPL}) was used.

We can see that our parametrization is a better fit to the three sets of observational data than a constant equation of state or a CPL parametrization since the value for $\chi^2$ is bigger for the latter than for the model introduced in \ref{eq:eos}.

\begin{table}	
	\begin{center}
	\begin{tabular}{|c|c|c|c|c|c|c|c|c|c|}
	\hline
	Alias & $\chi^2$ & $\chi_{red}^2$& $\nu$=9-M& $w_0$ &$w_i$ \\ \hline 
	 w constant & 3.61 &0.45 & 8 &-0.92 &- \\
	 \hline
	 CPL  & 3.29 & 0.47 & 7 &-0.98 &-0.59\\ \hline
	\end{tabular}
	\caption{Results from the minimization when the EoS is assumed to be  either a constant or given by the CPL parametrization using the  9 points of Table \ref{table:datarbao}. $\chi^2$ represents the minimum of the chi-squared function, $\chi_{red}^2=\chi^2/\nu$ the reduced value  and $w_0$, $w_i$ the parameters of the EoS for the DE.}
	\la{table:w-cpl-9pts}
	\end{center}
\end{table}

\begin{table}	
	\begin{center}
	\begin{tabular}{|c|c|c|c|c|c|c|c|c|}
	\hline
	Alias & $\chi^2$ & $\chi_{red}^2$& $\nu$=7-M& $w_0$ &$w_i$  \\ \hline
	w constant &  2.89 & 0.48  & 6 & -0.93 &- \\
	 \hline
	 CPL  & 2.87 & 0.57 & 5 &-0.96 &-0.70 \\
	 \hline
	\end{tabular}
	\caption{Results  for  the  7 points of Table \ref{table:datarbao} that exclude Ly$\alpha$-F measurements when the EoS was assumed to be either a constant, $w$, or given by the CPL parametrization.}
	\la{table:w-cpl-7pts}
	\end{center}
\end{table}

\begin{table}	
	\begin{center}
	\begin{tabular}{|c|c|c|c|c|c|c|c|c|}
	\hline
	Alias & $\chi^2$ & $\chi_{red}^2$& $\nu$=7-M& $w_0$ &$w_i$  \\ \hline
	w constant & 2.62 & 0.43 & 6 &-1.00 & -  \\
	 \hline
	  CPL  & 2.63 & 0.52 & 5 &-1.09 &-0.97  \\ 
	\hline
	
	\end{tabular}
	\caption{Results from the numerical minimization when  the  $D_V(z)$ measurements (Table \ref{table:datadvz}) were used and  the EoS was assumed to be either a constant, $w$, or given by the CPL parametrization.}
	\la{table:w-cpl-dvz}
	\end{center}
\end{table}


\section{Conclusions}

\la{sec:Conclusions}

We presented a parametrization for the equation of state of the Dark Energy and found the constraints deduced by using the BAO measurements contained in Tables \ref{table:datarbao} ($r_{BAO}(z)$ ) and \ref{table:datadvz} ($D_V(z)$). The  parametrization introduced (\ref{eq:eos}) includes the  widely used CPL  parameterization (\ci{Chevallier:2000qy, Linder:2002et}) as a particular case when $z_t = q = 1$, but it allows for a richer physical behavior. In par\-ti\-cu\-lar, this parametrization represents a fluid which performs a transition from a high redshift value $w_i$ to its present value $w_0$ at a given epoch, denoted by $z_t$ and modulated by the value of the exponent $q$.

As we have already mentioned, our main interested lied in studying the low redshift regime where the DE component starts to dominate the expansion of the Universe and so we chose the BAO distance measurements. Given the importance of this regime, a large amount of incoming measurements using galaxy redshift surveys will be available in the future (\cite{Levi:2013gra,Abbott:2005bi,Dawson:2015wdb,2011arXiv1110.3193L,2011arXiv1110.3193L,2009arXiv0912.0201L}).

By looking at the values for the $\chi^2$ function reported in the three tables summarizing our results and from Figure \ref{fig:data&curves} as well, we conclude that our parametrization is a good fit to the observational data and the fitting is better that using a constant EoS, $w$,  a cosmological constant, $\Lambda$, or the CPL parametrization.  

The values for the reduced chi-squared function, $\chi^2_{red}$, are smaller than unity for the ``$r_{BAO} 9$" results, in which case we see that the number of additional free parameters introduced is compensated with the number of data points available. 

In particular, when the ``$r_{BAO} 9$" set of measurements was used along  with Planck priors (outcome labeled as ``BAO+ Planck 1$\sigma$" in Table \ref{table:results9pts}) we found that an EoS with $w_0=-0.91$, $w_i=-0.62$ with a steep transition ($q$= 9.95) at $z_T=1.16$ was  the best fit to the data ($\chi^2$=2.85 and $\chi^2_{red}$=0.95).

When we use either the ``$r_{BAO} 7$" set or the ``$D_V(z)$" points,  the fit is still better than assuming $\Lambda$CDM, a constant EoS or the CPL parametrization (see tables \ref{table:results9pts}-\ref{table:resultsdvz} and \ref{table:w-cpl-9pts}-\ref{table:w-cpl-dvz}), but the extra free parameters are not compensated by the number of data points.

The scenario $q=z_T=1$ was included with $1\sigma$ of statistical significance in the following models:  the ``BAO + Planck" and ``BAO + Planck 1$\s$" from Table \ref{table:results9pts}, ``BAO$_{\Omega_m}$" and ``BAO + Planck$_{\Omega_m}$" from Table \ref{table:results7pts} and ``BAO + Planck" from Table \ref{table:resultsdvz}. 
 
We found smaller confidence contours, i.e., tighter constraints, for the  parametrization \ref{eq:eos}  (labeled as ``$w^q$" for brevity) than when we assumed  $\Lambda$CDM in the $h-\Omega_m$ space, as can be seen in figures \ref{fig:contoursLCDM}.  The use of BAO data alone prefers a higher $\Omega_m$ and lower $h$ value, as shown in Figure \ref{fig:contoursPlanck}, when compared to the priors from Planck. 

To summarize, the study of dynamics of Dark Energy is a matter of profounds implications for our understanding of the Universe and its physical laws.  Although the measurements from CMB polarization  are  the most precise data sets in Cosmology, the best way to analyze the properties  of DE comes from the low redshift regime, where the BAO feature is the most robust cosmic ruler. In this work  we have contributed towards that direction,  and we have presented the constraints for a dynamical DE model coming from the analysis of BAO distance measurements in the low redshift regime with and without Planck priors.

\acknowledgments
We acknowledge financial support from  DGAPA-PAPIIT IN101415  and Conacyt Fronteras de la Ciencia 281 projects. MJ thanks Conacyt for financial support.

\newpage
\section*{Appendix}

\subsection*{Growth of matter perturbations with background Dark Energy}

For the perturbative analysis with background Dark Energy we solved numerically the following differential equation:

\begin{equation}
a^2\delta_m''+a\frac{3}{2}\left(1-w(a)\Omega_{DE}(a)\right)\delta_m'- \frac{3}{2}\Omega_m(a)\delta_m=0
 \label{eq:growth}
\end{equation}

where $\delta_m(a)$ represents the matter density contrast and $w(a)$ is the parametrization for DE introduced in \ref{eq:eos} written in terms of the scale factor. We are taking zero perturbations for the DE component, i.e., $\delta_{DE}=0$. The initial conditions were  $\delta_{m,ini}=10^{-4}$ and $\delta'_{m,ini}=\frac{\delta_{m,ini}}{a_{ini}}$ at $a_{ini}=10^{-3}$, when the matter component dominates and  for a corresponding  wavenumber  of $k\approx 4\times10^{-3}$.

For the models chosen, we have set the value for $\Omega_m(a_0$) = 0.3156 $\equiv \overline{\Omega}_m$, as reported by Planck.

In  column (a) of Figure \ref{fig:perturbations} we display the growth function normalize to its initial value, $D_{+}(a)\equiv\frac{\delta_m(a)}{\delta_m(a_{ini})}$. We took a cosmological constant and the parametrization \ref{eq:eos} for $w(a)$ in \ref{eq:growth}, in particular, we used the best models reported as ``BAO + Planck" and ``BAO + Planck 1$\s$"   in tables \ref{table:results9pts}, \ref{table:results7pts}, and \ref{table:resultsdvz}. 

The bottom panel  of every plot shows the relative difference from each model to $\Lambda$CDM, $\Delta D_{+}(z)\equiv (D_{+}(z)-D_{+,\Lambda}(z))/D_{+,\Lambda}$(z), where $D_{+,\Lambda}(z)$ co\-rres\-ponds to the growth function when assuming a cosmological constant. From $\Delta_{+}$ we can see differences of  at most 5$\%$ (for the ``$r_{BAO}7$" data set).

In the right column of Figure \ref{fig:perturbations} we show the growth index, $\gamma(a) = \frac{f(a)}{\Omega_m(a)}$, where we assumed the well known ansatz $f(a)= \Omega_m^{\gamma}$ and  we compare the evolution of $\gamma(a)$ to its behavior when a cosmological constant is assumed.

This results show us that, even when considered only at a background level, the presence of DE modifies the evolution of the matter perturbations and a deeper analysis will be subject of a next paper \cite{Macorra:jaber2016perturbs}.


\clearpage

\bibliography{SteepEoS-Bibliography}
\bibliographystyle{ieeetr}

\end{document}